\documentclass[11pt]{article}
\pdfoutput=1
\usepackage{amssymb,amsmath,amsfonts,mathtools}
\usepackage{commath}
\usepackage{subcaption}
\usepackage{enumerate}
\usepackage{bbm,fp}
\usepackage{epsfig}
\usepackage{wrapfig}
\usepackage{yfonts}
\graphicspath{{./figures/}}	
\usepackage{arydshln}
\usepackage{geometry}                		% See geometry.pdf to learn the layout options. There are lots.
\usepackage{graphicx}				% Use pdf, png, jpg, or eps§ with pdflatex; use eps in DVI mode
\usepackage{xcolor}
\definecolor{hyperref}{RGB}{026,028,185}
\usepackage[bookmarks=true,colorlinks=true,linkcolor=hyperref,citecolor=hyperref,urlcolor=hyperref,bookmarksnumbered,hyperindex=true]{hyperref}
\usepackage{verbatim}
\usepackage{physics} %adds some practical stuff
\usepackage[sort,compress]{cite}
\numberwithin{equation}{section}
\renewcommand{\[}{\begin{equation}}
\renewcommand{\]}{\end{equation}}
\newcommand{\be}{\begin{eqnarray}}
\newcommand{\ee}{\end{eqnarray}}
\newcommand{\nn}{\nonumber}

\def\Z{ {\mathbb Z} }

\def\la{\lambda}
\def\G{\Gamma}

\def\half{{1\over 2}}

\def\D{\mathcal{D}}
\def\M{\mathcal{M}}

\newcommand{\B}{\mathcal{B}}

\def\tr{\mathrm{Tr}\hspace{1pt}}
\newcommand{\p}{\partial}

\newcommand{\nod}{n_o}
\newcommand{\nev}{n_e}

\newcommand{\cl}{\text{cl}}
\newcommand{\barphi}{\bar{\phi}}

\newcommand{\tildephi}{\tilde{\phi}}

\usepackage{color}
\definecolor{colorgreen}{HTML}{009688} % green

\newcommand{\diff}[1]{\dif^{\,#1}\!}

\def\lb{\label}
\def\la{\langle }
\def\ra{\rangle }
\def\DM{\partial\mathcal{M}}
\def\E{\mathcal{E}}
\def\O{\mathcal{O}}
\def\H{\mathcal{H}}
\def\eps{\epsilon}
\def\spec{{\rm Spectrum}}
\def\WE{W_{\hspace{-1pt}\E}}
\def\hp{\hspace{1pt}}

\begin{document}
\renewcommand{\thefootnote}{\arabic{footnote}}
 
\overfullrule=0pt
\parskip=2pt
\parindent=12pt
\headheight=0in \headsep=0in \topmargin=0in \oddsidemargin=0in

\vspace{ -3cm} \thispagestyle{empty} \vspace{-1cm}
\begin{flushright} 
\footnotesize
%PRE-PRINT number\\
\end{flushright}%

\begin{center}
\vspace{1.2cm}
{\Large\bf \mathversion{bold}
Logarithmic Negativity in Quantum Lifshitz Theories}

\vspace{0.8cm} {J.~Angel-Ramelli$^{\,a,}$\footnote{{\tt jfa1@hi.is}}, C. Berthiere$^{\,b, }$\footnote{{\tt clement.berthiere@pku.edu.cn}}, 
V.~Giangreco~M. Puletti$^{\,a,}$\footnote{{\tt vgmp@hi.is}}, and L.~Thorlacius$^{\,a,}$\footnote{{\tt lth@hi.is}}}

\vskip  0.5cm

\small
{\em
(a) University of Iceland,
Science Institute,
Dunhaga 3,  107 Reykjav\'ik, Iceland
\vskip 0.25cm
(b) Department of Physics, Peking University, Beijing 100871, China
}
\normalsize

 \end{center}

\vspace{0.3cm}
\begin{abstract}
We investigate quantum entanglement in a non-relativistic critical system by calculating the logarithmic negativity of a class of mixed states in the quantum Lifshitz model in one and two spatial dimensions.  
In 1+1 dimensions we employ a correlator approach to obtain analytic results for both open and periodic biharmonic chains. In 2+1 dimensions we use a replica method and consider spherical and toroidal spatial manifolds. 
In all cases, the universal finite part of the logarithmic negativity vanishes for mixed states defined on two disjoint components. 
For mixed states defined on adjacent components, we find a non-trivial logarithmic negativity reminiscent of two-dimensional conformal 
field theories. As a byproduct of our calculations, we obtain exact results for the odd entanglement entropy in 2+1 dimensions.
\end{abstract}
\newpage

\tableofcontents
\pagenumbering{arabic}

\setcounter{footnote}{0}
\newpage

%%%%%%%%%%%%%%%%%%%%%%%%%%%%%%%%%%%%%%%
%%%%%%%%%%%%%%%%%%%%%%%%%%%%%%%%%%%%%%%

\section{Introduction}

Over the past decade, quantum information theory has led to important insights and advances in several areas of physics including quantum field theory, condensed matter physics, and quantum gravity. Central to these developments is the concept of quantum entanglement, which constitutes a fundamental characteristic distinguishing quantum systems from classical ones. 
Quantum entanglement can be characterised in different ways and there is no single measure that captures all aspects of entanglement for all quantum systems. One particular measure, that has been the focus of numerous studies, is the entanglement entropy associated with a state described by a density matrix $\rho$, and a subsystem $A$ of the full system $A\cup B$. It is defined as $S_A=-\tr(\rho_A\log \rho_A)$, where the reduced density matrix is $\rho_A=\Tr_B\rho$ and the Hilbert space of the full system is assumed to factorise, $\H=\H_A\otimes\H_B$. This last assumption fails in practice for systems described by local quantum field theories, and this is manifested by short distance divergences appearing in the entanglement entropy. The leading divergence is usually a power law in the UV cutoff with a coefficient proportional to the area of the entangling surface that separates the subsystems. In the case of quantum critical theories, the entanglement entropy typically has a sub-leading logarithmic divergence with a scheme independent {\it universal\/} coefficient that encodes information about long-range entanglement in the system. In the present paper we will consider a class of quantum critical theories and focus our attention exclusively on  the universal terms. 

The entanglement entropy is particularly useful when the full system is in a pure quantum state, but this is rather restrictive.
In practice, one often has limited information about the system in question and needs to work with mixed quantum states. 
Thermal states are archetypal examples of such states, and, in this case, the entanglement entropy is no longer a good measure 
of quantum entanglement in the sense that it includes contributions from both quantum and classical correlations.
Other ways of quantifying entanglement besides entanglement entropy should then be introduced -- and they are legion~\cite{Plenio:2007zz, Amico:2008aa, Horodecki:2009aa}. 
A few important representatives are the entanglement cost and distillable entanglement \cite{Bennett:1995tk}, the entanglement of formation \cite{Bennett:1996gf}, and the logarithmic negativity \cite{Eisert:1998pz,Vidal:2002zz}.
In the crowded field of measures of bipartite entanglement for mixed states, the logarithmic negativity stands out as being actually computable. Indeed, most of the other entanglement measures, including the aforementioned, involve a minimisation 
over infinitely many quantum states, thus rendering them extremely difficult (if not impossible) to evaluate analytically in a quantum field theory setting. 

In the present work we consider the quantum entanglement of mixed states in a class of critical quantum field theories. 
For technical reasons we focus on mixed states that are simple to construct starting from a pure state but still reflect the 
essential issues arising for generic mixed states. Beginning with a system in a pure state, we take two non-overlapping 
subsystems, $A_1$ and $A_2$, that are not complements of one another ({\it i.e.} their complement defines a third subsystem $B$)
and consider the reduced density matrix on $A=A_1\cup A_2$, which is in general that of a mixed state. 
In a finite-dimensional Hilbert space, any mixed state can be purified by viewing it as a reduction of a pure state in a larger Hilbert space.
In a quantum field theory the corresponding question is more subtle, but the states we consider are purified by construction.

The logarithmic negativity introduced in \cite{Vidal:2002zz} is defined as follows. Let $\rho_A$ be the density matrix of a bipartite system $A=A_1\cup A_2$ in a pure or mixed state. We further suppose that the Hilbert space corresponding to our system factorises as $\H=\H_{A_1}\otimes\H_{A_2}$, and define $\vert e^{(1)}_i \ra$ and $\vert e^{(2)}_i \ra$ to be orthonormal basis states of $\H_{A_1}$ and $\H_{A_2}$, respectively, such that their tensor products $\vert e^{(1)}_i \ra \otimes\vert e^{(2)}_j \ra \equiv \vert e^{(1)}_i e^{(2)}_j \ra $ form a basis of $\H$. The partial transposition of the density matrix, with respect to $\H_{A_2}$, is an operator $\rho_A^{T_2}$ acting on $\H_{A_1}\otimes\H_{A_2}$ with matrix elements in the $\vert e^{(1)}_i  e^{(2)}_j\ra$ basis given by
\be
\la e^{(1)}_i e^{(2)}_j\vert\,\rho_A^{T_2}\,\vert e^{(1)}_k e^{(2)}_l\ra \equiv \la e^{(1)}_i e^{(2)}_l\vert\,\rho_A\,\vert e^{(1)}_k e^{(2)}_j\ra\,.\lb{PTdef}
\ee
In other words, the matrix elements of $\rho_A^{T_2}$ are obtained from those of $\rho_A$ by simply swapping basis elements $\vert e^{(2)}_j\ra\leftrightarrow \vert e^{(2)}_l\ra$ in $\H_{A_2}$.
Then the logarithmic negativity is obtained as
\be
\label{def-LN-general}
\E = \log ||\rho_A^{T_2}||\,,
\ee
where the trace norm $||\O||\equiv \tr\sqrt{\O^\dagger\O}$ is the sum of the absolute values of the eigenvalues of the operator $\O$. 
Its relevance relies on a crucial observation~\cite{Peres:1996dw,HORODECKI19961}: 
A necessary condition for the separability of the density matrix $\rho_A$ (that is for the system to be in a non-entangled state) is that its partial transpose (as e.g. $\rho_A^{T_2}$) is also a density matrix, which means that its spectrum is non-negative.
Partial transposition is not a unitary operation, and a non-vanishing $\E$ in \eqref{def-LN-general} detects when the system fails to be separable.
The logarithmic negativity, despite not being convex, is an entanglement monotone \cite{Plenio:2005cwa}, both under local quantum operations and classical communication (LOCC) and under positive partial transpose preserving operations (PPT). It is also additive and provides bounds on certain other measures \cite{Vidal:2002zz}. For pure states, the logarithmic negativity does not reduce to the entanglement entropy but instead coincides with the \mbox{R\'enyi entropy of order $1/2$.}

A replica method was developed in \cite{Calabrese:2012ew,Calabrese2013} for calculating the logarithmic negativity in many-body systems. 
In essence, the replica method relates the negativity to the traces of integer powers of $\rho_A^{T_2}$. Since the eigenvalues of $\rho_A^{T_2}$ are not guaranteed to be positive, 
the traces $\tr\big(\rho^{T_{2}}_A\big)^n$ are sensitive to the parity of $n$. Denoting by $n_e$ ($n_o$) the even (odd) integers, we obtain the 
 trace norm by analytic continuation of the even sequence at $n_e\rightarrow 1$, so that the logarithmic negativity reads
\be
\E=\lim_{n_e\rightarrow 1} \log\tr\big(\rho^{T_{2}}_A\big)^{n_e}\,. \lb{LNreplica}
\ee
This approach has been extensively applied to ground states in conformal field theory (CFT) \cite{Calabrese:2012ew,Calabrese2013,Calabrese:2013mi,DeNobili:2015dla,Coser:2015eba,Shapourian:2016cqu}, but also at finite temperature \cite{Calabrese:2014yza,Shapourian:2018lsz} and in out-of-equilibrium situations \cite{Coser:2014gsa,Hoogeveen:2014bqa,Wen:2015qwa}, as well as to topological systems \cite{Wen:2016snr,Wen:2016bla}. 

In this paper, we are interested in a certain class of non-relativistic quantum field theories -- those admitting Lifshitz symmetry. Lifshitz field theories exhibit anisotropic scaling between space and time,
\be
t\rightarrow \lambda^z\,  t\,,\qquad \mathbf{x}\rightarrow\lambda\, \mathbf{x}\,,
\ee
with characteristic dynamical exponent $z > 1$. 
Non-relativistic theories are especially relevant in the context of condensed matter physics. In particular, the Lifshitz theory in $2+1$ dimensions with dynamical exponent $z=2$, referred to as the quantum Lifshitz model (QLM)~\cite{Ardonne2004}, is known to describe a quantum phase transition in systems, such as quantum dimer models \cite{Ardonne2004,PhysRevLett.61.2376}, between a uniform phase and a phase with spontaneously broken translation invariance in two spatial dimensions. 
The $(2+1)$-dimensional QLM was generalised to $d+1$ dimensions with a critical exponent $z=d$ in \cite{Keranen2017} where this special class of Lifshitz theories was named generalised quantum Lifshitz models (GQLMs).
A key feature of these ($d+1$)-dimensional Lifshitz field theories with (even) positive integer $z$ is that the ground state wave-functional takes a local form, given in terms of the action of a $d$-dimensional Euclidean CFT. 
The local nature of the ground state makes these theories rare examples of non-relativistic theories which admit analytic treatment.
The entanglement properties of ground states of quantum Lifshitz theories have been extensively studied \cite{Fradkin:2006mb,PhysRevB.80.184421,Hsu:2008af,Hsu:2010ag,Oshikawa:2010kv,Zaletel2011,Zhou:2016ykv,Angel-Ramelli:2019nji,Berthiere:2019lks} using analytic and numerical methods. 

In the present paper, we extend the work on entanglement in Lifshitz field theories by evaluating analytically the logarithmic negativity for a class of bipartite mixed states in the quantum Lifshitz model. The mixed states are obtained by tracing out the degrees of freedom of one of the subsystems in a tripartite pure state, which for us will be the ground state of the QLM.
We adopt two different approaches to the calculation of logarithmic negativity. 
First, we employ the so-called correlator method~\cite{Audenaert:2002xfl,2003JPhA...36L.205P}, which in essence discretises the theory on a lattice. For this part we consider the $(1+1)$-dimensional version of the theory with Lifshitz exponent $z=2$. 
We then compute the logarithmic negativity by means of the replica method~\cite{Calabrese:2012ew,Calabrese2013}, with focus on the $(2+1)$-dimensional QLM defined on two different spatial manifolds, a 2--sphere and a 2--torus. 

In both approaches, we start the discussion by considering a bipartite system in its ground state and confirm that in this case the logarithmic negativity reduces to the $n=1/2$ R\'enyi entropy, as it should for a system in a pure state~\cite{Calabrese:2012ew,Calabrese2013}. 
After that, we investigate a system in a more general mixed state, obtained by partially tracing over the ground state of a bipartite system, resulting in a reduced density matrix $\rho_A$.
We then further divide the subsystem  $A$ into $A_1$ and $A_2$ and partially transpose over $A_2$ in order to compute the logarithmic negativity. At this point, we analyse two different cases, depending on whether the subsystems $A_1$ and $A_2$ are disjoint or adjacent when viewed as part of the original tripartite system. 

Interestingly, the logarithmic negativity turns out to vanish for disjoint subsystems in the QLM, both in one and two spatial dimensions.
This is unexpected in a gapless system and is in sharp contrast with 2d CFT~\cite{Calabrese:2012ew,Calabrese2013}. The physical 
origin of this effect is not clear to us but it is a robust result that we obtain using two different approaches: a correlator method for a 
discrete theory in one spatial dimension and a replica method for a continuum theory in two spatial dimensions. 
In the discrete non-compact theory, the reduced density matrix on disjoint intervals for the open chain is separable%
\footnote{Perhaps this has its origin in the local nature of the ground state.},
which is in general a stronger result than the vanishing of the logarithmic negativity. 
It remains an open question whether the corresponding reduced density matrix for disjoint submanifolds, obtained via the 
replica method is also separable. Similar results were found previously for the topological logarithmic negativity in Chern-Simons 
theory~\cite{Wen:2016snr,Wen:2016bla}, as well as in a $(1 + 1)$-dimensional system with $z=2$ Lifshitz scaling~\cite{Chen:2017txi},
that is closely related to our discrete theory.
The resemblance between QLM and topological theories was first noted in \cite{Fradkin:2006mb,Hsu:2008af,Zhou:2016ykv}, where the entanglement entropy for the QLM was found to exhibit a finite sub-leading universal term analogous to the topological entanglement entropy.
%It was, however, also noted by those authors that these finite terms have different physical origins in the QLM compared to topological theories. 
For adjacent subsystems we obtain a non-trivial logarithmic negativity, which is somewhat closer to $2d$ CFT results~\cite{Calabrese:2012ew,Calabrese2013}. 

A numerical study of logarithmic negativity in Lifshitz theories in one and two spatial dimensions for arbitrary $z$ was carried out in~\cite{MohammadiMozaffar:2017chk}. Our findings partially confirm their results, but we emphasise that our approach is entirely analytical. By concentrating on the QLM with $z=2$, we are able to obtain closed form expressions for the logarithmic negativity, both in the correlator approach and the replica method. As far as we know, this is the first time the replica method is used to compute the logarithmic negativity in Lifshitz theories, and, for the discrete theory in one spatial dimension, we have obtained moments of the $z=2$ QLM reduced density matrix and its partial transpose in analytic form -- something that is still beyond reach for the relativistic boson ($z=1$). 
In the present work, we have chosen to focus on the special case of $z=2$ and $d=1$ or $2$, but several of our results generalise to other values of $z$ and $d$ and we comment on this along the way.

As a by-product of our study we also obtain the so-called odd entanglement entropy, or odd entropy for short, in the $(2+1)$-dimensional QLM. The main motivation for considering the odd entropy is to have an entanglement measure that directly computes the entanglement wedge cross section in holographic two-dimensional CFTs~\cite{Tamaoka:2018ned}. 

The paper is organised as follows.
In Section \ref{sec:QLM} we briefly review key definitions for the QLM. 
In Section \ref{sec:LN-correlator} we obtain the logarithmic negativity in a $(1+1)$-dimensional model by means of the correlator method. 
We then proceed in Section  \ref{sec:LN-replica-method} to calculate the logarithmic negativity via a replica method using path integrals. 
Our results on odd entropy are presented in Section \ref{sec:odd-EE} and in Section \ref{sec:discussion} we conclude with a discussion 
and some open questions. 
Some technical details related to the correlator method appear in Appendix \ref{Apdx1}, and details related to the replica approach are 
found in Appendices \ref{app:det} and \ref{app:winding-torus}. Appendix \ref{app:OE} completes Section \ref{sec:odd-EE} on odd entropy, and this work.

%%%%%%%%%%%%%%%%%%%%%%%%%%%%%%%%%%%%%%%%%
%%%%%%%%%%%%%%%%%%%%%%%%%%%%%%%%%%%%%%%%%

\section{Logarithmic negativity from correlation functions}
\label{sec:LN-correlator}

%%%%%%%%%%%%%%%%%%%%%%%%%%%%%%%%%%%%%%%%%
\subsection{The quantum Lifshitz model}
\label{sec:QLM}

The $(2+1)$-dimensional quantum Lifshitz model, with critical exponent $z=2$ on the spatial manifold $\M$, is a quantum field theory involving a compact scalar field $\phi\sim\phi+2\pi R_c$ defined by the Hamiltonian \cite{Ardonne2004}
\begin{equation}
\label{hamiltonian-d=z=2}
H=\frac{1}{2}\int_{\M}\diff{2}x \left(\pi^2+g^2 (\triangle \phi)^2\right),
\end{equation}
where $\pi=-i\delta/\delta\phi$ is the momentum conjugate to the field, $\triangle$ is the Laplacian on $\M$, and $g$ is a free parameter of the model. The ground state can be expressed in terms of a path integral of a two-dimensional Euclidean theory~\cite{Ardonne2004},
\begin{align}
\label{ground-state}
\vert \Psi_0\rangle &= \frac{1}{\sqrt{Z_\M}} \int\D\phi\ e^{-\frac{1}{2}S[\phi]}\vert\phi\rangle, \hspace{-.5cm}& S[\phi]&=g\int_\M\diff{2}x\ (\nabla\phi)^2,
\end{align} 
with the partition function given by $Z_\M:=\int\D\phi\ e^{-S[\phi]}$. We denote the corresponding density matrix by 
\begin{equation}
\label{ground-state-density-matrix-d=z=2}
\rho:=\vert\Psi_0\rangle\langle \Psi_0 \vert=\frac{1}{Z_\M}\int\D\phi\D\phi'\, e^{-\frac{1}{2}(S[\phi]+S[\phi'])}\vert\phi\rangle\langle\phi'\vert.
\end{equation}

A $1+1$-dimensional quantum Lifshitz model with $z=2$ can be defined in analogous fashion, with the Laplacian replaced by $\partial_x^2$ and the integration measure by $\dif x$. We will take the scalar field to be \textit{non-compact} in the $1+1$-dimensional case.

Generalisations to higher spatial dimensions $d$ and even integer critical exponents $z$ are possible, with some restrictions~\cite{Keranen2017,Angel-Ramelli:2019nji}. For instance, when the spatial manifold is a $d$--sphere the even critical exponent $z$ is required to satisfy $z\le d$ in order to guarantee a well-defined GJMS-operator~\cite{Angel-Ramelli:2019nji,gjms:1992}. Generalizations to higher odd integer values of $z$ are less understood and will not be considered here.
Since the physically relevant systems are in one and two spatial dimensions, we will restrict our calculations to $d=1$ and $d=2$, but point out whenever our results are valid beyond those cases.

%%%%%%%%%%%%%%%%%%%%%%%%%%%%%%%%%%%%%%%%%%%%%%%%%%%%%%%%%%%%%%%%%%%%%%%%%%%%%%%%%%%%%%%%%%%%%%%%%%%%%%%%%%%%%%%%%%%%%%%%%%%%%%%%%%%%%%%%%%

\subsection{Logarithmic negativity from correlator method} 

The correlator method for computing the entanglement entropy or logarithmic negativity of Gaussian states has a long tradition \cite{Audenaert:2002xfl,2003JPhA...36L.205P,Casini:2006hu,Casini:2009sr,Calabrese2013,PhysRevB.93.115148,DeNobili:2016nmj,Helmes:2016fcp,Berthiere:2018ouo,MohammadiMozaffar:2017nri,MohammadiMozaffar:2017chk}. This method has been almost exclusively employed as a numerical one, often as a check of field theory predictions. Here, we focus on the $(1+1)$-dimensional free Lifshitz scalar field with dynamical exponent $z=2$ in its ground state. We obtain simple closed form results for the R\'enyi entropies and logarithmic negativity for the discrete theory on a one-dimensional lattice, which are then easily translated to the continuum.
\begin{figure}[h]
\centering
\vspace{10pt}
\includegraphics[scale=1.]{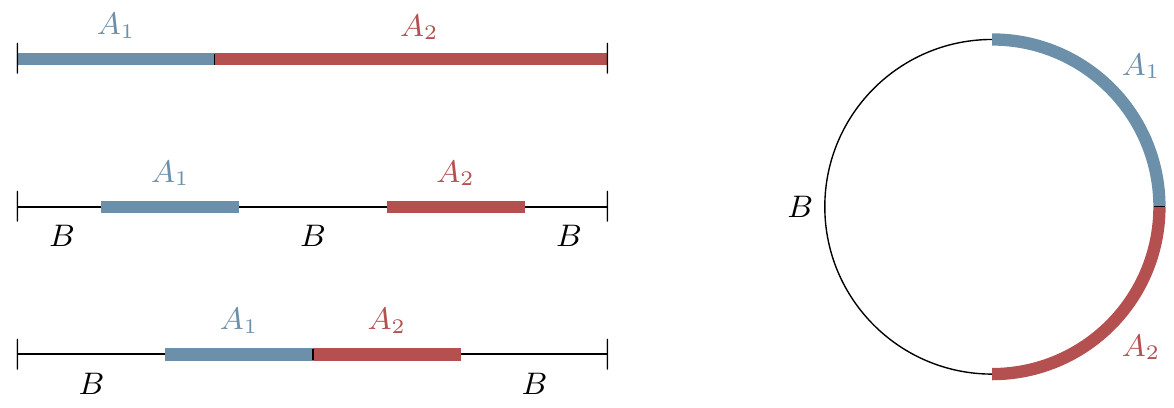}
%\vspace{0pt}
\caption{Entanglement between two intervals $A_1$ and $A_2$ embedded in the ground state of a (larger) system formed by the union of $A_1$, $A_2$ and the complement $B$. Left: open system with Dirichlet boundary conditions at both ends. Right: Periodic system.} 
\lb{intervals}
\end{figure}

\paragraph{Discrete theory and boundary conditions.}
The Hamiltonian of a non-compact free massless scalar field $\phi$ with dynamical exponent $z=2$ in $1+1$ dimensions is given by
 \be
H= \frac12 \int_\M dx \left(\pi^2 + \phi \,\partial^4_x\,\phi \right),\lb{Hcont}
 \ee
where $\M$ is the one-dimensional line with open or periodic boundary condition. Without loss of generality, we have set to unity the constant $g$ appearing in front of the spatial derivatives. Discretising the theory on a lattice with $L$ sites, the above Hamiltonian is replaced by
 \be
 H= \frac12 \left(\pi^T\pi + \phi^T K\phi \right), \lb{Hdiscrete}
 \ee
where $\phi^T=(\phi_1 ,\phi_2,\cdots, \phi_L)$, $\pi^T=(\pi_1 ,\pi_2, \cdots, \pi_L)$, and the matrix $K$ is a discrete version of the spatial biharmonic operator $\triangle^2\equiv\partial^4_x$. Static solutions of the Hamiltonians \eqref{Hcont} and \eqref{Hdiscrete} satisfy 
\be
\triangle^2\phi=0\,, \qquad {\rm and}\qquad K\phi=0\,,\lb{eomsD}
\ee 
respectively, with some specified boundary conditions at the boundary $\DM$ of the space/lattice. The biharmonic equation requires additional boundary conditions compared to the standard Laplace equation. In the continuum theory a natural ``Dirichlet" boundary condition is given by
 \be
\phi |_{_{\DM}}=0\,, \quad{\rm and}\quad \triangle\phi |_{_{\DM}}=0 \,.\lb{DBC1}
 \ee
A lattice version of this Dirichlet condition can be implemented as follows. First, introduce degrees of freedom on fictitious lattice sites at the boundaries, $\phi_{-1}$, $\phi_0$, $\phi_{L+1}$, $\phi_{L+2}$ and impose the lattice biharmonic equation of motion,
\be
\phi_{i-2}-4\phi_{i-1} + 6\phi_i -4\phi_{i+1} +\phi_{i+2}= 0\,,\quad\lb{eoms}
\ee
for $i=1,\ldots L$, where we have set the lattice spacing $\epsilon$ to unity for simplicity. The fictitious fields appear in the equations for $\phi_1$, $\phi_2$, $\phi_{L-1}$ and $\phi_L$ but they can be eliminated by imposing a discrete version of the Dirichlet conditions $\phi |_{_{\DM}}=0$ and $\triangle\phi |_{_{\DM}}=0$, given by
\be
\phi_0=0\,,\quad{\rm and}\quad 
-\phi_{-1}+2\phi_0-\phi_{1}=0\,,
\ee
at $i=0$, and similarly at $i=L+1$. With these boundary conditions, the matrix $K$ is indeed simply the square of the discrete Laplacian matrix with standard Dirichlet boundary conditions. % and reads
% \be
% K=
% \left(\begin{array}{ccccc}
% 5 & -4 & 1 &  & \\
% -4 & 6 & -4 & \ddots & \\
%  1& -4 & \ddots & \ddots & 1 \\
%  & \ddots &\ddots & 6 & -4 \\
%   &  & 1  & -4 & 5
% \end{array}\right).\quad
% \ee
%
Alternatively, one can impose periodic boundary conditions on the lattice. The resulting $K$ is the square of the usual discrete Laplacian matrix with periodic boundary conditions. Note, however, that the matrix $K$ has a vanishing eigenvalue for a periodic chain and is non-invertible unless a mass term is added to the Hamiltonian \eqref{Hdiscrete}.

\paragraph{Correlation functions, reduced density matrix and partial transpose.}

Vacuum two-point functions are given by
\be
X_{ij}\equiv \la \phi_i\phi_j \ra = \frac{1}{2}(K^{-1/2})_{ij}\,, \quad{\rm and}\quad P_{ij}\equiv\la \pi_{i}\pi_j \ra = \frac{1}{2}{(K^{1/2})}_{ij}\,.\lb{corr}
\ee
The reduced density matrix $\rho_A$ can easily be related \cite{2003JPhA...36L.205P} to the correlation matrices $X$ and $P$ restricted to the subsystem $A$ (denoted hereafter $X_A$ and $P_A$). In particular, from the eigenvalues $\{\nu_i\}_{i=1,\cdots,\ell}$ of $C_A=\sqrt{X_AP_A}$ for a region $A$ of size $\ell$, the trace of the $n^{\rm th}$ power of the reduced density matrix $\rho_A$ reads
\be
\tr\rho_A^n=\prod_{i=1}^{\ell} \bigg[\Big(\nu_i+\frac{1}{2}\Big)^n -\Big(\nu_i-\frac{1}{2}\Big)^n \bigg]^{-1},\quad\lb{RE}
\ee
from which one easily obtains the R\'enyi entropies
\be
S_A^{(n)}=\frac{1}{1-n}\log\tr\rho_A^n\,. \lb{REE}
\ee

Now consider a tripartite system with $A= A_1\cup A_2$.
The partial transposition with respect to, e.g., $A_2$, for a bosonic Gaussian state, corresponds to time reversal applied only on the momenta corresponding to the subsystem $A_2$ \cite{Audenaert:2002xfl}. The partially transposed reduced density matrix $\rho_A^{T_2}$ thus remains a Gaussian matrix. We introduce the matrices
\be
P_{A}^{T_2}&=&T_2\cdot P_{A} \cdot T_2\,,\\
T_2&=&\mathbbm{1}_{\ell_1}\oplus(-\mathbbm{1}_{\ell_2})\,,
\ee
where $\ell_1,\,\ell_2$ are the lengths of the intervals $A_1,\,A_2$, respectively, such that $\ell=\ell_1+\ell_2$. The trace of the $n^{\rm th}$ power of $\rho_{A}^{T_2}$ can then be computed from the eigenvalues $\{\lambda_i\}_{i=1,\cdots,\ell}$ of $C_{A}^{T_2}\equiv\sqrt{X_{A}P_{A}^{T_2}}$ as
\be
\tr\big(\rho_A^{T_2}\big)^n=\prod_{i=1}^{\ell} \bigg[\Big(\lambda_i+\frac{1}{2}\Big)^n -\Big(\lambda_i-\frac{1}{2}\Big)^n \bigg]^{-1},\lb{rhoPT}
\ee
from which the trace norm follows straightforwardly  
\be
||\rho_{A}^{T_2}|| = \prod_{i=1}^{\ell}\bigg[\Big|\lambda_i+\frac{1}{2}\Big| -\Big|\lambda_i-\frac{1}{2}\Big| \bigg]^{-1}= \prod_{i=1}^{\ell}\max\left(1,\frac{1}{2\lambda_i}\right).
\ee
Finally, the logarithmic negativity is given by
\be
\E = \sum_{i=1}^{\ell}\log\left[\max\left(1,\frac{1}{2\lambda_i}\right)\right].
\ee
Notice that only the eigenvalues that satisfy $\lambda_i<1/2$ contribute to the logarithmic negativity.

\subsection{R\'enyi entropies}
We start by computing R\'enyi entropies for a single interval in a bipartite pure state. This allows us to carry out a simple 
consistency check of our calculations in the discrete model by evaluating the logarithmic negativity for the same interval
and confirming that it reduces to the R\'enyi entropy of order $n=1/2$, as it should for a pure state.

\paragraph{Open system.}
For a finite chain of $L$ lattice sites with Dirichlet boundary conditions at both ends, the vacuum two-point functions \eqref{corr} take a simple form,
\be
X_{ij} &=& \frac{1}{2(L+1)}
\begin{cases}
i(L-j+1)\,, & i\le j \vspace{3pt}\\
j(L-i+1)\,, & i>j
\end{cases}\\
P_{ij} &=& \delta_{ij} - \frac12(\delta_{i,j-1}+\delta_{i-1,j})\,. \lb{Pcorr}
\ee
Taking an interval $A_1=[1,\ell]$ adjacent to one of the boundaries ($A_2=\bar{A}_1$, see \mbox{Fig.~\ref{intervals}} top-left panel), one finds that the matrix $C_{A_1}$ is triangular with spectrum
\be
\spec(C_{A_1}) = \left\{\frac{1}{2}\sqrt{\frac{(\ell+1)(L-\ell+1)}{L+1}},\, \frac{1}{2},\,\cdots,\, \frac{1}{2} \right\}.\lb{ev1}
\ee
Quite remarkably, only one eigenvalue, $\nu\equiv\nu_1$, contributes to the entropy. Plugging $\nu$ in \eqref{REE} yields exact expressions for the R\'enyi entropies.

We can access the continuum regime of the theory by reintroducing the lattice spacing $\epsilon$ into the notation via $L\rightarrow L/\eps$ and $\ell\rightarrow \ell/\eps$, and taking the limit $\eps\rightarrow0$. The continuum R\'enyi entropies read
\be
S_{A_1}^{(n)} = \frac{1}{2}\log(\frac{\ell(L-\ell)}{\eps\hspace{1pt}L}) +\frac{1}{n-1}\log\big(2^{1-n}n\big)  \,. \lb{EElat1}
\ee
The leading term in \eqref{EElat1} is independent of the R\'enyi index $n$ as expected \cite{Fradkin:2006mb}, and agrees with the results of \cite{Chen:2017txi,Chen:2017tij} where the Renyi entropies were obtained by mapping the ground state of the $z=2$ boson to that of a path integral for a quantum mechanical particle. The finite part is non universal and depends on how one regulates the theory in the UV.

%%%%%%%%%%%%%%%%%%%%%%%%%%%
\paragraph{Periodic system.}
For periodic boundary conditions, the $K$ matrix of a finite biharmonic chain is a circulant matrix. It is non-invertible due to a zero eigenvalue but the zero mode can be lifted by adding a mass term, ${1\over 2} m^4\phi^2$, to the Hamiltonian \eqref{Hcont} resulting in \be
K={\rm circ}(6 + m^4,-4,1 ,0,\, \cdots,0,1,-4)\,.
\ee 
The mass $m$ has dimensions of inverse length and is measured in units of the inverse lattice spacing $\epsilon^{-1}$, which has been set to one as before. 

We are interested in the critical regime, that is $m\rightarrow0$ and $m L\ll1$, for which the eigenvalues of $C_{A_1}$ constructed from the circulant matrix $K$ on a single interval $A_1$ of size $\ell$ reduce to
\be
\spec(C_{A_1}) = \left\{\frac{1}{\sqrt{2m^2L}},\, \frac{1}{2}\sqrt{\frac{(\ell+1)(L-\ell+1)}{2L}},\, \frac{1}2,\,\cdots,\,\frac{1}2  \right\}.\lb{ev2}
\ee
If we reinstate the lattice spacing $\epsilon$ and take 
the continuum limit as before, we obtain the following expression for the single interval R\'enyi entropies,
\be\label{pureRenyii}
S_{A_1}^{(n)} =  \frac{1}{2}\log(\frac{\ell(L-\ell)}{\eps\hspace{1pt} L}) -\frac{1}{2}\log{(\eps m^2L)} +\frac{2}{n-1}\log\big(2^{1-n}n\big)+ \cdots \,,
\ee
where the ellipsis denotes terms vanishing in the limits $m L\ll 1$, $\eps\rightarrow0$.

\subsection{Logarithmic negativity}\lb{sec:LNcorr}
Let us now turn to the logarithmic negativity. We have to compute the eigenvalues of the matrix $C_A^{T_2}$, defined above $\eqref{rhoPT}$, for a bipartite (sub)system $A=A_1\cup A_2$ of the $z=2$ chain. We first consider the pure state case, for which $A$ is the whole system, then we move on to the configuration of two disjoint intervals $A_1$ and $A_2$, and finally we let $A_1$ and $A_2$ be adjacent.

\subsubsection{Pure states}

When $\rho_{A}$ is pure, $A_2$ is the complement of $A_1$, and $\ell_1=L-\ell_2$, see \mbox{Fig.~\ref{intervals}} top-left panel. In that case, for the open chain, there is only one eigenvalue of $C_{A}^{T_2}$ that satisfies $\lambda_i<1/2$, that is
\be
\lambda^2 %&=& \frac{1}4 - \frac{L_{A}(L-L_{A})}{2(L+1)}\left(\sqrt{\frac{(L_{A}+1)(L-L_{A}+1)}{L_{A}(L-L_{A})}}-1 \right).\\
&=& 2\nu\Big(\nu -\sqrt{\nu^2 - 1/4}\Big) -1/4\,,
\ee
where $\nu$ is the single eigenvalue of $C_{A_1}$ in \eqref{ev1}, with $\ell\equiv\ell_1$, that is distinct from $1/2$. After some algebra, one obtains that the logarithmic negativity of this bipartite pure state is, as expected, the R\'enyi entropy of order $1/2$, i.e. $\E(\rho_{A_1\cup A_2}) = S_{A_1}^{(1/2)}$.

Similarly, for the periodic chain in the critical regime, only two eigenvalues of $C_{A}^{T_2}$ contribute to the logarithmic negativity, i.e.
\be
\lambda^2_1 = \frac{m^2L}{32}\,,\qquad \lambda^2_2 =  2\nu\Big(\nu -\sqrt{\nu^2 - 1/4}\Big) -1/4\,,
\ee
where $\nu$ is the second eigenvalue of $C_{A_1}$ in \eqref{ev2}, with $\ell\equiv\ell_1$. One can then check that for the periodic case as well, the logarithmic negativity reduces to the $(1/2)$--R\'enyi entropy of $A_1$.

\subsubsection{Two disjoint intervals}
\label{sec:two-disjoint-intervals}
Now let $A$ be a subsystem of the $z=2$ open chain of length $L$ and further divide $A$ into two subsystems, $A=A_1\cup A_2$ with $A_1$ and $A_2$ disjoint, as for example depicted in \mbox{Fig.~\ref{intervals}} middle-left panel.
First, take $A_1$ and $A_2$ to be of the same size $\ell_1=\ell_2=\ell$ and each adjacent to one of the boundaries of the total system. The distance between $A_1$ and $A_2$ is then $d=L-2\ell>0$. We find in that case,
\be
\spec(C_A^{T_2}) = \left\{\frac{\sqrt{\ell+1}}{2},\, \frac{1}{2}\sqrt{\frac{(\ell+1)(L-2\ell+1)}{L+1}},\, \frac{1}2,\,\cdots,\,\frac{1}2  \right\}.\lb{ev3}
\ee
A quick inspection of the spectrum \eqref{ev3} of $C_A^{T_2}$ reveals that not a single eigenvalue is smaller than $1/2$. We thus conclude that the logarithmic negativity vanishes, $\E=0$, for this configuration of two disjoint intervals. 
This result may seem surprising. It is, however, a consequence of the separability of the reduced density matrix. 
To arrive at that conclusion, we rely on the following statement proven in \cite{Lami_2018}: A bipartite non-compact Gaussian state 
that is invariant under partial transposition of one of the two subsystems is separable. It is easy to see that for two disjoint regions in the $z=2$ chain, the corresponding reduced density matrix is indeed invariant under partial transposition, cf. Appendix \ref{Apdx}, and thus separable. It follows that the logarithmic negativity is zero.

A vanishing logarithmic negativity on disjoint intervals was observed previously in a closely related $(1+1)$-dimensional system with Lifshitz scaling in \cite{Chen:2017txi}. These authors study the ground state of the Motzkin Hamiltonian subject to the constraint $\phi\ge0$, which renders the density matrix non-Gaussian. However, for two intervals far away from the boundaries of the system, the constraint becomes unimportant and the model reduces to the $z=2$ free boson studied in the present paper. In contrast to \cite{Chen:2017txi}, our result applies regardless of whether the two disjoint regions are located near or far away from the boundaries of the system, 
and also on a circle of finite length. 
We will see below that the same result is found in the $(2+1)$-dimensional quantum Lifshitz model and extends to higher-dimensional models with Lifshitz scaling as well.

In a slightly more general case, where $A_1$ and $A_2$ are symmetric with respect to the center of the chain, but not necessarily adjacent to the boundaries and separated by a distance $d>0$, the eigenvalues of $C_{A}^{T_2}$ distinct from $1/2$ are the (positive) solutions of the following two equations:
\be
 32 \lambda^4 - 8(L - \ell-d+2) \lambda^2 + (\ell+1) (L -2 \ell-d+2)&=&0\,,\\
 32(L+1) \lambda^4 - 8(L+2-d^2+(\ell+d)(L - 2\ell+1)) \lambda^2 &&\nonumber\\
 + (\ell+1)(d+1) (L -2 \ell-d+2)&=&0\,.
\ee
As before, the UV cutoff can be restored by making the changes $L\rightarrow L/\eps$, $\ell\rightarrow \ell/\eps$ and $d\rightarrow d/\eps$. For both $L$ and $\ell$ arbitrary, the two solutions of the first equation above are always larger or equal to $1/2$, while for the second equation one finds that its solutions may be smaller than $1/2$, but only provided $d<\eps$. However, since the UV cutoff $\eps$ is arbitrarily small in the continuum regime, neither of the eigenvalues can actually be smaller than $1/2$, thus implying, again that
$\E = 0$.
More generally, we find that the logarithmic negativity vanishes for arbitrary configurations of two disjoint intervals. This may also easily be verified numerically. The same conclusion carries through to the periodic chain. 

In \cite{MohammadiMozaffar:2017chk}, it was observed based on numerical computations that for high values of the dynamical exponent $z$, there exist a critical distance between two disjoint intervals below which the logarithmic negativity is non-vanishing. We believe this to be a lattice effect. In our analytic calculation above, we found that in the continuum regime and upon restoring the UV cutoff $\eps$, the critical distance is actually proportional to $\eps$. We saw this explicitly for $z=2$, but it is also true for $z>2$ where the critical distance can be shown to be $d_c=(z/2)\eps$. Later on we will see, using path integrals and the replica method, that the logarithmic negativity vanishes for two disjoint systems for any even positive integer $z$.

%%%%%%%%%%%%%%%%%%%%%%%%%%%
\subsubsection{Two adjacent intervals}

\paragraph{Open system.}

Now consider two intervals of same length $\ell$ joined at the center of the full system 
(assuming $L$ even), as shown in the bottom-left panel of \mbox{Fig.~\ref{intervals}}. In this case, the only eigenvalue of $C_{A}^{T_2}$ satisfying $\lambda<1/2$ reads
\be
\lambda^2 = \frac{(2\ell+1)(L-2\ell+2)+2}{16(L+1)}\left(1-\sqrt{1-\frac{8(L+1)(L-2\ell+2)}{((2\ell+1)(L-2\ell+2)+2)^2}} \right).
\ee
In the continuum regime, with all lengths measured in units of the UV cutoff $\eps$ from now on, we have $\lambda^{-1}=\sqrt{8\ell}$, from which follows the logarithmic negativity
\be
\E = \frac{1}{2}\log(2\ell)\,. \lb{LNA1}
\ee
Notice that for $\ell=L/2$, the negativity \eqref{LNA1} reduces to the $(1/2)$--R\'enyi entropy \eqref{EElat1}. Indeed, in that case $\rho_A$ is pure.

In the most general case, that is for two adjacent intervals of arbitrary lengths and relative position in the total system, there are at most four eigenvalues of $C_{A}^{T_2}$ distinct from $1/2$. These four eigenvalues are the roots of a certain quartic equation presented in Appendix \ref{Apdx1}. What is important here is that only one eigenvalue, call it $\lambda$, among these four roots is smaller than $1/2$, and we find in the continuum regime that
\be
\lambda= \sqrt{\frac{\ell_1+\ell_2}{16\ell_1\ell_2}}\,.
\ee
The logarithmic negativity of two adjacent intervals in a finite system with Dirichlet boundary conditions is thus given in general by
\be
\E = \frac{1}{2}\log(\frac{\ell_1 \ell_2}{\ell_1+\ell_2}) + const\,, \lb{LNA2}
\ee
where $const = \log2$ in our setup here, but is not a universal quantity and depends on the regularisation scheme. For $\ell_1=\ell_2=\ell$ we recover \eqref{LNA1}.

\paragraph{Periodic system.}
Let us now consider a finite system of length $L$ with periodic boundary conditions, and two adjacent intervals of lengths $\ell_1$ and $\ell_2$ such that $\ell_1+\ell_2\le L$, as in the right-hand panel of \mbox{Fig.~\ref{intervals}}. As discussed above, the discrete theory has a divergence due to a zero mode that we circumvent by introducing a non-zero mass. Working in the limit of very small mass, one might expect a term logarithmic in the mass parameter to appear in the negativity, as is indeed the case for pure states with $\ell_1+\ell_2=L$ where the logarithmic negativity equals the $(1/2)$--R\'enyi entropy given by \eqref{pureRenyii}. 
It turns out, however, for a mixed state such that $\ell_1+\ell_2<L$, no divergent term appears in the logarithmic negativity.
In the simplest case where $\ell_1=\ell_2=\ell < L/2$, the spectrum of $C_{A}^{T_2}$ in the continuum limit is found to be
\be\label{ct2_spectrum}
\spec(C_A^{T_2}) = \left\{\sqrt{\frac{1}{8\ell}},\sqrt{\frac{\ell}{6}},\sqrt{\frac{\ell(L-2\ell)}{8L}},\sqrt{\frac{3}{2m^2L}},\frac{1}{2},\,\cdots,\frac{1}{2}\right\}.
\ee
In the critical regime, where $m L\ll 1$, the only eigenvalue that contributes to the logarithmic negativity is $\lambda_1=\sqrt{1/(8\ell)}$
and we get $\E=(1/2)\log(2\ell)$, the same as for the open chain with Dirichlet boundary conditions. Note that this result is only reliable 
for a mixed state where the strict inequality $\ell < L/2$ holds. Indeed, for $\ell=L/2$, the third eigenvalue in the expression 
\eqref{ct2_spectrum} for the spectrum vanishes, indicating that the regulator mass needs to be retained and in this case the 
logarithmic mass dependence of the pure state result \eqref{pureRenyii} is recovered.

In the general case, with arbitrary $\ell_1+\ell_2<L$, the spectrum of $C_{A}^{T_2}$ in the critical limit is given in Appendix \ref{Apdx1}. The spectrum contains only one eigenvalue smaller than $1/2$, which in the continuum regime reads $\lambda^2=(\ell_1+\ell_2)/(16\ell_1\ell_2)$, and we find the same logarithmic negativity as for the open system.

%%%%%%%%%%%%%%%%%%%%%%%%%%
\subsubsection{A hint at a general formula}
\label{sec:hint-at-a-general-formula}

Let us first emphasise that, for the $z=2$ free boson, we find the expression 
\be
\E = \frac{1}{2}\log(\frac{\ell_1\ell_2}{\ell_1+\ell_2}) + const \lb{LNgen}
\ee
for the continuum logarithmic negativity of two adjacent intervals in a finite or infinite system, with or without (Dirichlet) boundaries. This is in  contrast to the $z=1$ relativistic free scalar field for which the logarithmic negativity of two adjacent intervals depends in general on the size of the total system as, e.g. for periodic boundary conditions
\be
\E^{(z=1)} = \frac{1}{4}\log\hspace{-2pt}\Bigg(\frac{L}{\pi}\frac{\sin\big(\frac{\pi \ell_1}{L}\big)\sin\big(\frac{\pi \ell_2}{L}\big)}{\sin\hspace{-2pt}\big(\frac{\pi(\ell_1+\ell_2)}{L}\big)}\Bigg)+ const\,,\lb{LNrelat}
\ee
and only for an infinite system $L\rightarrow\infty$ one obtains
\be
\E^{(z=1)} = \frac{1}{4}\log(\frac{\ell_1\ell_2}{\ell_1+\ell_2}) + const \,. 
\ee
Since a picture is worth a thousand words, we plot in Fig.~\ref{plot1} the logarithmic negativities of two adjacent intervals of same length $\ell$ in a periodic chain of length $L$ for a $z=2$ and a $z=1$ scalar. One can appreciate the difference in behaviour between the two theories, particularly close to $\ell\simeq L/2$ where $\E^{(z=2)} \propto \log L$ while $\E^{(z=1)} \propto \log(L^2/(L-2\ell))$.

\begin{figure}[h]
\centering
\vspace{10pt}
\includegraphics[scale=1.1]{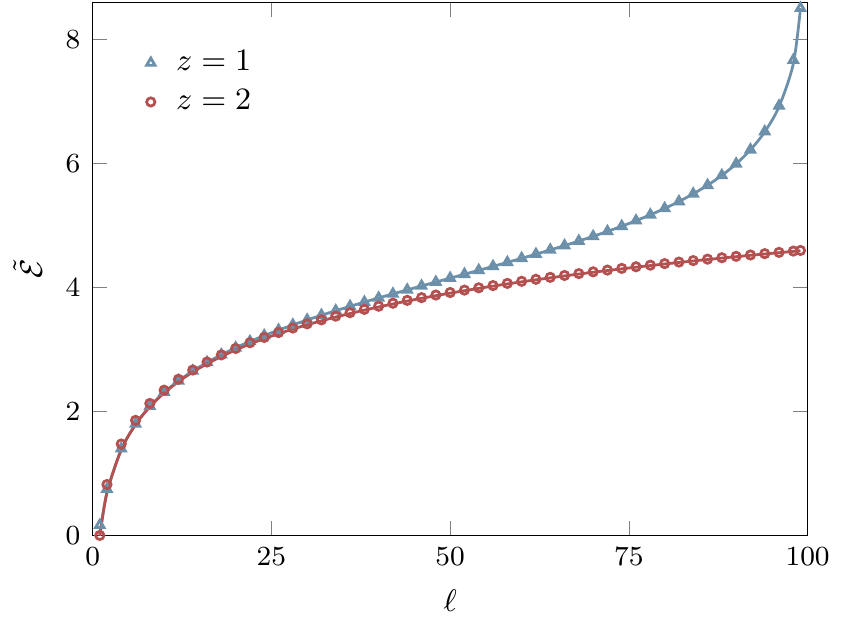}
\vspace{-5pt}
\caption{Logarithmic negativities of two adjacent intervals of same length $\ell$ in the periodic chain of length $L=200$ and mass $m=10^{-5}$ for the relativistic ($z=1$) and Lifshitz ($z=2$) bosons. To allow an easy comparison between the two theories, the logarithmic negativities are normalised in such a way that for $\ell\ll L$ they behave as $\tilde{\E}\simeq\log\ell$. The data are perfectly consistent with the (normalised) continuum expressions \eqref{LNgen} and \eqref{LNrelat}, shown as solid lines.} 
\lb{plot1}
\vspace{10pt}
\end{figure}

Fradkin and Moore \cite{Fradkin:2006mb} taught us that the bipartite R\'enyi entropies for ground states of non-compact scalar fields with critical dynamical exponent $z=2$ can be simply expressed in terms of partition functions of a free Euclidean CFT in one dimension lower, namely
\be
S^{(n)}_A = -\log(\frac{Z_{A} Z_{B}}{Z_{A\cup B}}) \,, \lb{fradk} 
\ee
and is actually independent of the R\'enyi index $n$. $Z_{A}$ and $Z_{B}$ are the CFT partition functions on regions $A$ and $B$, respectively, with Dirichlet boundary conditions on the entangling cut. $Z_{A\cup B}$ is the partition function on the entire space, with specified boundary conditions, for example Dirichlet, at the boundary $\DM$. Returning to the logarithmic negativity for mixed states of two adjacent intervals, we have found that the negativity \eqref{LNgen} does not depend on the size of the total system. Furthermore, we know that for a pure state it reduces to the R\'enyi entropy of order $n=1/2$, which for the $z=2$ scalar is given by \eqref{fradk} independently of $n$.
We are thus led to conjecture the following general formula for the logarithmic negativity of the $z=2$ non-compact free scalar field:
\be
\E = -\log(\frac{Z_{A_1} Z_{A_2}}{Z_{A_1\cup A_2}}) \,, \lb{LNconj} 
\ee
where $Z_{A_i}$ is the partition function of the Euclidean CFT in one dimension lower on $A_i$ with Dirichlet boundary conditions on the entangling cut(s), and $Z_{A_1\cup A_2}$ is the partition function on $A_1\cup A_2$ with similar boundary conditions. 
Clearly, when $A_2$ is the complement of $A_1$, formula \eqref{LNconj} reduces to the entropy \eqref{fradk}. When $A_1$ and $A_2$ are disjoint, these regions do not talk to each other because of the Dirichlet boundary conditions, thus one has  $Z_{A_1\cup A_2} = Z_{A_1}Z_{A_2}$, hence $\E=0$. Finally, if $A_1$ and $A_2$ are adjacent, using heat kernel techniques one can easily compute in $1d$ (omitting non-universal parts): $-\log Z_{A_{1,2}}=(1/2)\log \ell_{1,2}$ and $-\log Z_{A_1\cup A_2}=(1/2)\log(\ell_1+\ell_2)$, such that we recover \eqref{LNgen}.

In the following section, we show that \eqref{LNconj} is indeed correct in the $(2+1)$-dimensional quantum Lifshitz model. It also holds for non-compact $(d+1)$-dimensional Lifshitz theories with even exponent $z$ on flat space and up to some subtleties on curved manifolds as well. We derive \eqref{LNconj} and its generalisation to compact fields using path integrals and the replica trick.

%%%%%%%%%%%%%%%%%%%%%%%%%%%%%%%%%%%%%%%%%%%%%%%%%%%%%%%%%%%%%%%%%%%%%%%%%%%%%%%%%%%%%%%%%%%%%%%%%%%%%%%%%%%%%%%%%%%%%%%%%%%%%%%%%%%%%%%%%%%%%%%%%%%%%%%%%%%%%%%%%%%%%%%%%%%%%%%%%%

\section{Logarithmic negativity from a replica approach}
\label{sec:LN-replica-method}

In this section, we apply replica techniques to evaluate the logarithmic negativity in the (2+1)-dimensional QLM.
The calculation is closely patterned on~\cite{Fradkin:2006mb,Zaletel2011,Zhou:2016ykv}, where a replica 
method was developed to calculate the entanglement entropy in the QLM.
As its name suggests, this method introduces independent copies of the original theory -- the replicas -- 
and a surgery procedure to join them together. The crucial step is to identify the correct set of boundary 
conditions to be imposed at the entangling cuts on the replica fields. 
We will adapt the technique to evaluate the logarithmic negativity in quantum Lifshitz theories on different spatial manifolds 
by means of an expression of the form \eqref{LNreplica}, for mixed state density matrices constructed from the ground 
state by partially tracing over a subsystem. 
%

%%%%%%%%%%%%%%%%%%%%%%%%%%%%%%%%%
\subsection{Pure states}
\label{sec:logarithmic-negativity-of-pure-states}

We begin, as in Section~\ref{sec:LNcorr}, by considering the logarithmic negativity of pure states, which should 
reduce to the R\'enyi entropy of order $1/2$.
In this case, the spatial manifold $\M$ is divided into two submanifolds $A_1$ and $A_2$, with boundary $\Gamma$ between them, 
and we assume the Hilbert space on the full manifold factorises as $\mathcal{H}=\mathcal{H}_{A_1}\otimes\mathcal{H}_{A_2}$. 
We then introduce a replica index $i=1,\ldots,n_e$ for the density matrices and rewrite \eqref{ground-state-density-matrix-d=z=2} on the bipartite manifold as
\begin{multline}
\label{rho-i-1}
\rho_i=\frac{1}{Z_{A_1\cup A_2}}\int\D\phi_i^{A_1}\D\phi_i'^{A_1}\D\phi_i^{A_2}\D\phi_i'^{A_2}\, e^{-\frac{1}{2}(S[\phi_i^{A_1}]+S[\phi_i'^{A_1}]+S[\phi_i^{A_2}]+S[\phi_i'^{A_2}])}\\
\cross \vert\phi_i^{A_1}\rangle\otimes\vert\phi_i^{A_2}\rangle\langle\phi_i'^{A_1}\vert\otimes\langle\phi_i'^{A_2}\vert\,.
\end{multline}
Note that since the replicated fields are all dummy fields we have $\rho_i\equiv \rho$. 
We stress that in the replica method we always work with the action of a free conformal compactified bosonic field for $z=2$, that is the action $S$ appearing in the above density matrix is given by the expression \eqref{ground-state}. 
This means that for $z=2$ it is enough to impose Dirichlet boundary conditions on the fields to have a self-adjoint Laplacian operator. 
The partial transposition over, e.g., $A_2$, then amounts to exchanging the primed and \mbox{unprimed $A_2$-fields in \eqref{rho-i-1}:}
\begin{multline}
\label{rho-i-2}
\rho_i^{T_{2}}=\frac{1}{Z_{A_1\cup A_2}}\int\D\phi_i^{A_1}\D\phi_i'^{A_1}\D\phi_i^{A_2}\D\phi_i'^{A_2}\, e^{-\frac{1}{2}(S[\phi_i^{A_1}]+S[\phi_i'^{A_1}]+S[\phi_i^{A_2}]+S[\phi_i'^{A_2}])}\\
\cross \vert\phi_i^{A_1}\rangle\otimes\vert\phi_i'^{A_2}\rangle\langle\phi_i'^{A_1}\vert\otimes\langle\phi_i^{A_2}\vert\,.
\end{multline}
We can now compute the trace of the $n_e$-th power of the partial transpose density matrix, $\tr\big(\rho^{T_2}\big)^{n_e}\equiv\tr\big(\rho^{T_2}_1\cdots\rho_{n_e}^{T_2}\big)$.
For $i=1,\ldots,\nev-1$, each adjacent matrix product $\rho_i^{T_{2}}\rho_{i+1}^{T_{2}}$ leads to two $\delta$-functions coming from $\langle\phi_i'^{A_1}\vert\phi_{i+1}^{A_1}\rangle$ and $\langle\phi_i^{A_2}\vert\phi_{i+1}'^{A_2}\rangle$. The final total trace of the product of density matrices adds another two $\delta$-functions  $\langle\phi_n'^{A_1}\vert\phi_{1}^{A_1}\rangle$ and $\langle\phi_n^{A_2}\vert\phi_{1}'^{A_2}\rangle$.
Resolving all the $\delta$-functions leads to the gluing conditions 
\begin{equation}
\label{gluing-conditions-pure-state}
\left.
\begin{aligned}
\phi_{i}'^{A_1}&=\phi_{i+1}^{A_1}\\
\phi_{i}^{A_2}&=\phi_{i+1}'^{A_2}\; 
\end{aligned}
\right\rbrace\qquad i=1,\ldots,n_e\,,
\end{equation}
with $\phi_{n_e+1}\equiv\phi_1$. 
\begin{figure}[h]
\centering
\vspace{20pt}
\includegraphics[scale=1]{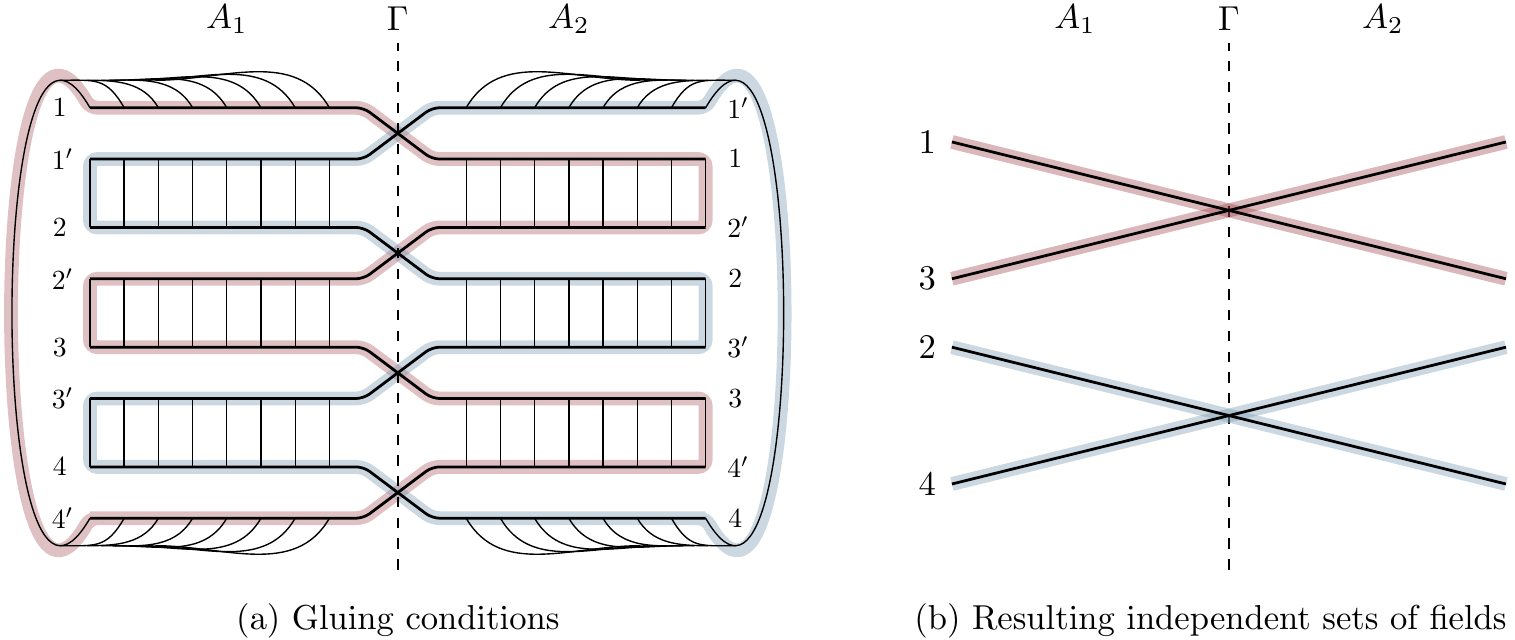}
\caption{Gluing conditions for $n_e=4$.  
Gluing results in two independent sets of boundary conditions represented in red and blue.}
\label{fig:boundary-conditions-pure-state}
\vspace{10pt}
\end{figure}
Furthermore, the continuity conditions among the fields at the entangling cut read 
\be
\phi^{A_1}_i\vert_\G =\phi^{A_2}_i\vert_\G \,, \qquad \quad {\phi'}^{A_1}_i\vert_\G ={\phi'}^{A_2}_i\vert_\G,
\ee
as can be seen in Fig.~\ref{fig:boundary-conditions-pure-state}. 
A closer look at these conditions reveals that all even and all odd fields must agree separately at the entangling cut $\Gamma$, leaving us with $n_e$ independent fields with boundary conditions 
\begin{equation}\lb{BCadjEO}
   \left.
   \begin{aligned}
   \phi^{A_1}_{2k}\vert_\Gamma &=\;\phi^{A_2}_{2l}\vert_\Gamma\hspace{4pt}\equiv&\hspace*{-6pt}\chi^e\;\\
   \phi^{A_1}_{2k-1}\vert_\Gamma &=\phi^{A_2}_{2l-1}\vert_\Gamma\equiv&\hspace*{-6pt}\chi^o\;
   \end{aligned}
   \right\rbrace\qquad k,l=1,\ldots, n_e/2,
\end{equation}
where $\chi^e$ and $\chi^o$ are two independent functions of the boundary coordinates. The partial transposition thus has the effect of creating two independent sets of $n_e/2$ fields. Since the boundary functions $\chi^{e,o}$ and the fields are all dummy integration variables, we can relabel them as $\phi^{A_i}_{2k},\phi^{A_i}_{2k-1}\mapsto \phi^{A_i}_{k}$ and $\chi^e,\chi^o\mapsto \chi$ to get
\begin{align}
\tr\big(\rho^{T_2}\big)^{n_e}
%&=\frac{1}{Z_{A_1\cup A_2}^{n_e}}\int_{\mathcal{B}}\prod_{i=1}^{n_e}\D\phi_i^{A_1}\, e^{-S[\phi_i^{A_1}]}\int_{\mathcal{B}}\prod_{i=1}^{n_e}\D\phi_i^{A_2}\, e^{-S[\phi_i^{A_2}]}\\
&=\left(\frac{1}{Z_{A_1\cup A_2}^{n_e/2}}\int_{\mathcal{B}}\prod_{k=1}^{n_e/2}\D\phi_k^{A_1}\, e^{-S[\phi_k^{A_1}]}\int_{\mathcal{B}}\prod_{k=1}^{n_e/2}\D\phi_k^{A_2}\, e^{-S[\phi_k^{A_2}]}\right)^2,
\label{rho-i-3}
\end{align}
where the boundary conditions $\mathcal{B}$ are now given by
\begin{equation}
   \mathcal{B}:\qquad \phi^{A_1}_k\vert_\Gamma=\phi^{A_2}_{l}\vert_\Gamma=\chi,\qquad k,l=1,\ldots,n_e/2\,.\lb{BCpure}
\end{equation}
Upon closer inspection, one can recognise in \eqref{rho-i-3} the expression for $\Tr \rho_{A_1}^{n_e/2}$ derived in \cite{Zaletel2011,Zhou:2016ykv,Angel-Ramelli:2019nji}, where $\rho_{A_1}=\Tr_{A_2}\rho$ is the reduced density matrix obtained by tracing out the degrees of freedom in $A_2$%
\footnote{Note that we could have written the expression in terms of $\rho_{A_2}=\Tr_{A_1}\rho$, since the system is in a pure state.},
meaning that the following equation holds
\be
\tr\big(\rho^{T_2}\big)^{n_e}= \Big(\tr \rho_{A_1}^{n_e/2}\Big)^2\,.
\ee 
In particular this gives $\lim_{n_e\rightarrow 2} \tr\big(\rho^{T_{2}}\big)^{n_e} = 1$, as it should \cite{Calabrese2013}.
For a compact field on a circle of radius $R_c$, the fields are subject to the boundary conditions \eqref{BCpure} up to the periodic identification $\phi\sim\phi+2\pi R_c$. In \cite{Zaletel2011,Zhou:2016ykv,Angel-Ramelli:2019nji} it was found that
\begin{equation}
\Tr\rho_{A_1}^{n_e/2}=\left(\frac{Z_{A_1}Z_{A_2}}{Z_{A_1\cup A_2}}\right)^{n_e/2-1}W(n_e/2)\,,
\end{equation}
where $Z_{A_i}$ is the partition function on $A_i$ with Dirichlet boundary conditions at the entangling cut $\Gamma$ and $W(n)$ is a sum over different classical configurations of the compactified fields. 
Applying the replica formula \eqref{LNreplica}, we obtain for the logarithmic negativity of a pure state
\begin{equation}
\label{logarithmic-negativity-pure-state}
\mathcal{E}=-\log(\frac{Z_{A_1}Z_{A_2}}{Z_{A_1\cup A_2}})+2\log W(1/2)\,
\end{equation}
which, as expected \cite{Calabrese2013}, is indeed the ($1/2$)--R\'enyi entropy $S^{(1/2)}_{A_1}$. 

It is also worth looking at the odd $n_o$ sequence $\tr\big(\rho^{T_2}\big)^{n_o}$. 
In that case, all the fields have to be equal at the entangling cut $\Gamma$, that is
\begin{equation}
\phi^{A_1}_i\vert_\Gamma=\phi^{A_2}_{j}\vert_\Gamma=\chi,\qquad i,j=1,\ldots,n_o\,.
\end{equation}
We thus have 
\be\label{nodd-pure}
\tr\big(\rho^{T_2}\big)^{n_o}= \tr \rho_{A_1}^{n_o}\,,
\ee 
which yields the normalization 
\be
\lim_{n_o\rightarrow 1} \tr\big(\rho^{T_{2}}\big)^{n_o} = W(1) = 1\,.
\ee

%%%%%%%%%%%%%%%%%%%%%%%%%%%%%%%%%%%%%%%%%%%
\subsection{Disjoint submanifolds}
\label{sec:disjointLN}
\begin{figure}[h]
\centering
\vspace{10pt}
\includegraphics[scale=1]{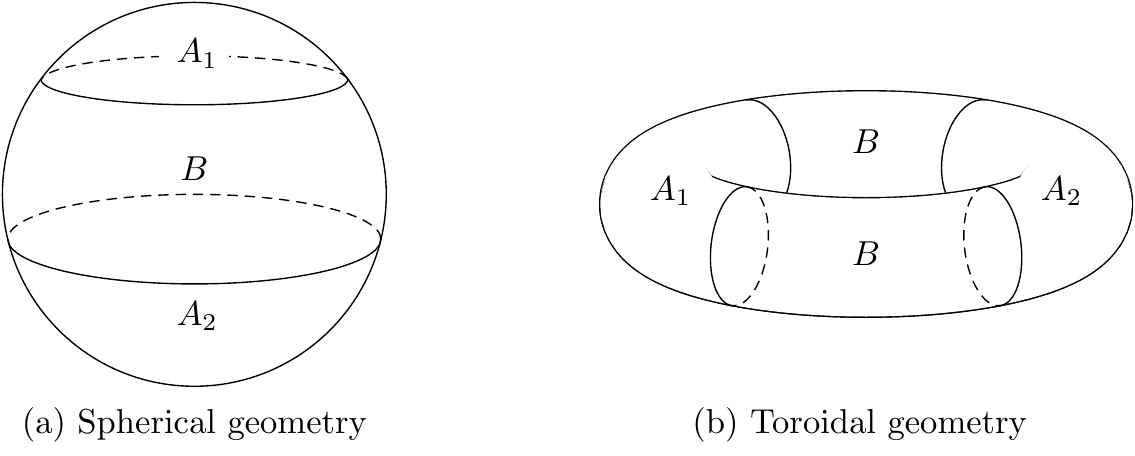}
\caption{Examples of geometries where $A_1$ and $A_2$ are separated by $B$. Note that for the torus, $B$ consists of two disjoint components.}
   \label{fig:realizations-of-disjoint-geometry}
\vspace{5pt}
\end{figure}

We now turn to the more interesting case of entanglement between two regions of a system in a mixed state.
In this section, we illustrate the replica approach for the case of a mixed state when $A_1$ and $A_2$ are disjoint and 
separated by $B$, as illustrated in Fig.~\ref{fig:realizations-of-disjoint-geometry}. 

\begin{figure}[h]
\centering
\vspace{10pt}
\includegraphics[scale=1]{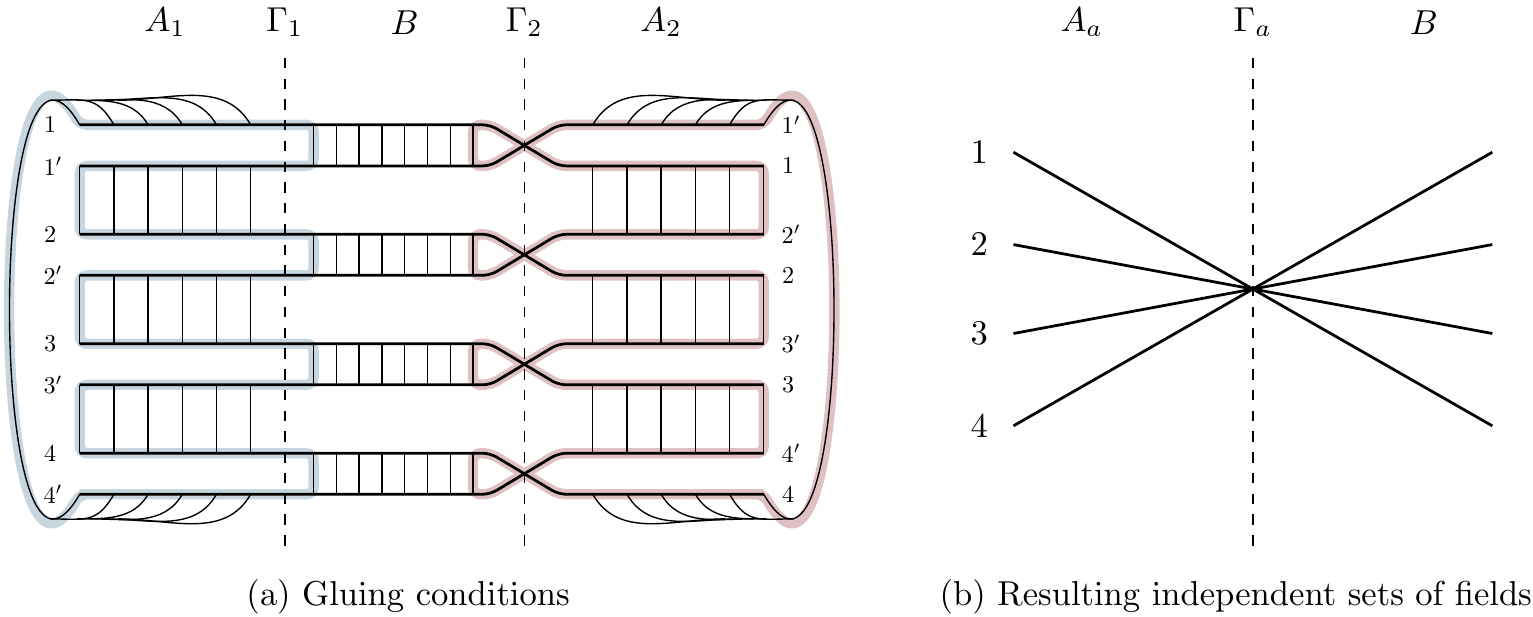}
   \caption{Gluing conditions around the boundaries between $B$ and the two components of $A$ for $n_e=4$. The resulting boundary conditions, depicted on the right, are the same for $A_1$ and $A_2$. }
   \label{fig:boundary-conditions-of-disjoint-geometry}
\end{figure}

The mixed state we consider is obtained by tracing over the degrees of freedom on $B$, 
with the full system in its ground state, and is thus described by the reduced density matrix $\rho_A\equiv \rho_{A_1\cup A_2}$. 
In order to calculate the logarithmic negativity, we then transpose the density matrix over $A_2$ resulting in $\rho_A^{T_2}$. 
The trace on $B$ leads to conditions of the form 
\begin{equation}
\label{gluing-conditions-B}
\phi^B_i=\phi'^B_i\,,
\end{equation}
that is the primed and unprimed copies of the fields are sewed within the same replica of the density matrix. The gluing conditions that result for the fields on $A_1$ and $A_2$ are the same as before, that is  \eqref{gluing-conditions-pure-state}, so they connect the density matrices cyclically.
The continuity conditions at the entangling cut between $A_a$ and $B$ (indicated as $\G_a$) require that 
$$\phi^{A_a}_i\vert_{\G_a} =\phi^{B}_i\vert_{\G_a} \,, \qquad {\phi'}^{A_a}_i\vert_{\G_a} ={\phi'}^{B}_i\vert_{\G_a}\,,$$
for all $i=1, \dots, \nev$ and $a=1,2$. 
Putting everything together, all replica fields must agree at the boundary between $B$ and any of the $A_a$'s, as depicted in Fig.~\ref{fig:boundary-conditions-of-disjoint-geometry}. In particular, this means that the geometry is \emph{not} sensitive to the partial transposition, and we obtain the identity
\begin{equation}
\label{rel-trace-rhot-general}
\tr\big(\rho_{A}^{T_{2}}\big)^{n_e}=\tr\rho_{A}^{n_e}\,.
\end{equation}
The latter quantity appears in the calculation of the tripartite entanglement entropy  \cite{Zhou:2016ykv} and is given by
\begin{equation}
\label{trace-rho-t-disjoint}
\Tr(\rho_{A})^{n_e}=\left(\frac{Z_{A_1}Z_{A_2}Z_{B}}{Z_{A\cup B}}\right)^{n_e-1}W(n_e).
\end{equation}
For disjoint subsystems the partial transposition is not sensitive to the parity of $n$, {\it i.e.} equations \eqref{rel-trace-rhot-general} and \eqref{trace-rho-t-disjoint} are also valid for odd $n_o$. It then immediately follows from the unit normalization of the density matrix in the odd sequence at $n_o=1$ that the winding sector contribution is trivial, $W(1)=1$.
We thus find the striking result,
\begin{equation}
\label{curlyE-disjoint-replica}
\E=\lim_{n_e\rightarrow 1}\log\tr\big(\rho_{A}^{T_{2}}\big)^{n_e}=0 \,.
\end{equation}
While this result differs from the expectation for a conformal field theory \cite{Calabrese2013}, it agrees with the correlator method calculations in Section~\ref{sec:two-disjoint-intervals}. 
We stress that the vanishing of the logarithmic negativity is a necessary but not sufficient condition for the separability of the density matrix~\cite{Peres:1996dw}. The theorem of \cite{Lami_2018} only applies to finite dimensional systems, so it remains an open question whether the reduced density matrix constructed via the replica method is separable for disjoint subsystems.

The expression \eqref{trace-rho-t-disjoint} also holds in higher dimensions for generalised quantum Lifshitz models with even $z$ as discussed in \cite{Keranen2017, Angel-Ramelli:2019nji}, as long as the cuts are smooth and a direct generalisation of Fig.~\ref{fig:realizations-of-disjoint-geometry}.
The relation \eqref{rel-trace-rhot-general} is therefore still valid for smooth partitions of the ground state of such theories, which implies that the main conclusion in \eqref{curlyE-disjoint-replica} remains correct. 
In the case of $d=z=2$ it suffices to impose Dirichlet boundary conditions on the fluctuations to ensure that the variational problem is well-posed and the Laplacian self-adjoint after surgery, leading to a consistent replica calculation. 
For curved higher-dimensional manifolds and higher $z$, further restrictions apply in order to have a well-defined higher-derivative operator in the action $S$ in \eqref{ground-state}.
On a $d$-sphere, for instance, the operator in question is only well-defined for even $z\le d$. For the construction of consistent operators on tori and spheres, and details on the corresponding replica calculation see \cite{Angel-Ramelli:2019nji} and references therein.

The fact that in Lifshitz theories with even dynamical exponent the entanglement negativity vanishes for disjoint subsystems is 
surprising but it is not unheard of. Similar behaviour was already noted in a closely related $z=2$ system in \cite{Chen:2017txi} and in
Chern-Simons field theories in $2+1$ dimensions the topological logarithmic negativity vanishes for disjoint subsystems \cite{Wen:2016snr,Wen:2016bla}. In this respect, Lifshitz theories exhibit similarities to topological theories.

%%%%%%%%%%%%%%%%%%%%%%%%%%%%%%
%%%%%%%%%%%%%%%%%%%%%%%%%%%%%%

\subsection{Adjacent submanifolds without winding}
\label{sec:adjacent-no-winding}

\begin{figure}[h]
\centering
\vspace{10pt}
\includegraphics[scale=1]{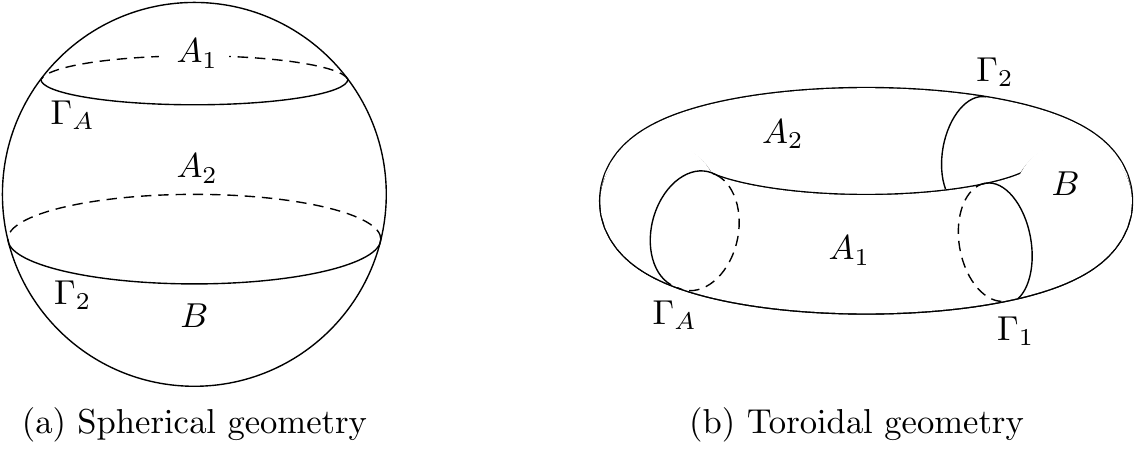}
\caption{Realisations of the situation when $A_1$ and $A_2$ are adjacent on the sphere and torus.}
   \label{fig:realizations-of-adjacent-geometry}
\end{figure}

Next, we consider the case where the submanifolds $A_1$ and $A_2$ are adjacent, as in Fig.~\ref{fig:realizations-of-adjacent-geometry}. 
To keep the discussion as general as possible, we assume the maximal number of non-trivial entangling cuts $\Gamma_1$, $\Gamma_2$, and $\Gamma_A$. The spherical case, which requires only two cuts, is recovered by trivially identifying fields across the third cut.
We take $\phi$ to be non-compact throughout this section and postpone addressing the additional complications that arise from 
the winding structure of a compact $\phi$ until Section \ref{sec:adjacent-manifolds-winding}.

\begin{figure}[h]
\centering
\vspace{10pt}
\includegraphics[scale=1]{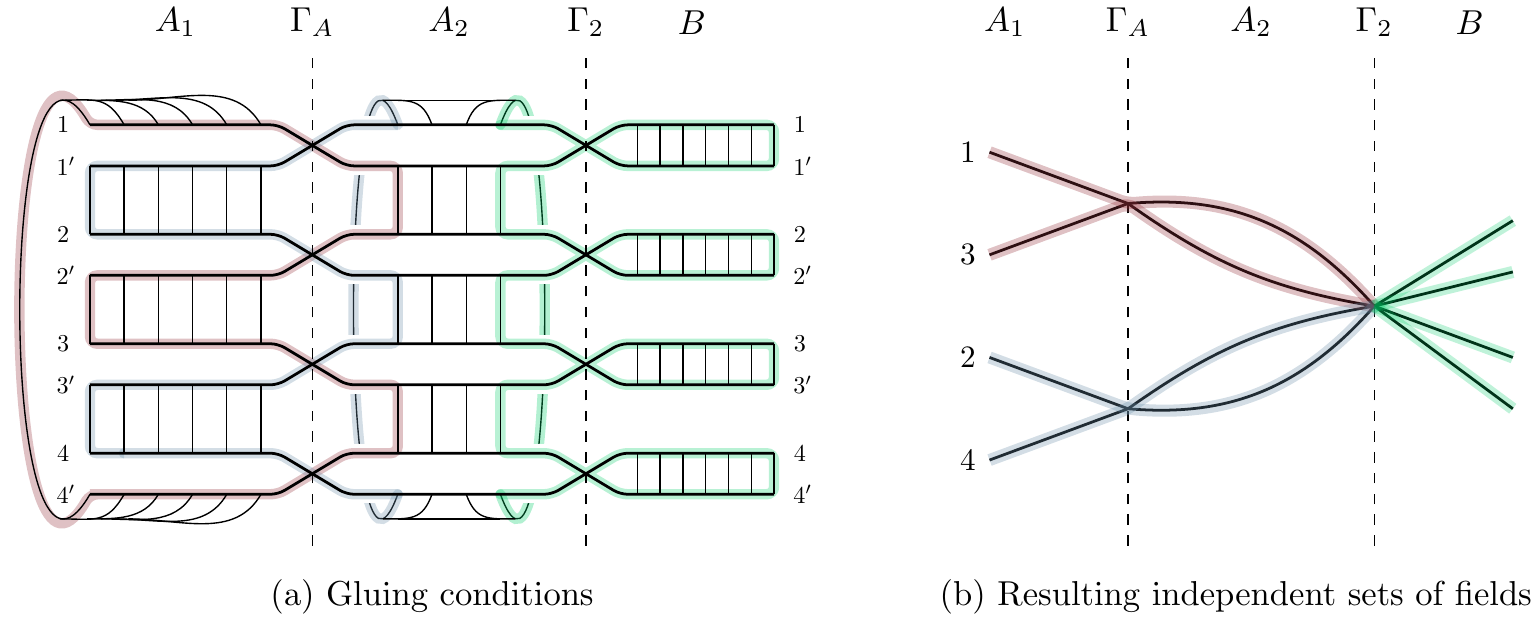}
\caption{Realisations ($\nev=4$) of the situation when $A_1$ and $A_2$ are adjacent on the sphere and torus. }
   \label{fig:gluing-adjacent}
\end{figure}

As in Section~\ref{sec:disjointLN}, we perform a trace over the degrees of freedom on $B$ at the beginning and then compute $\tr\big(\rho_{A}^{T_{2}}\big)^{n_e}$ with a transposition on $A_2$. 
The partial trace on B leads to the gluing conditions \eqref{gluing-conditions-B}, while the product and final trace over $A_1, A_2$ leads to the conditions \eqref{gluing-conditions-pure-state}. 
At the entangling cut between $A_1$ and $A_2$, denoted by $\G_A$, the continuity conditions are
\be
\phi^{A_1}_i\vert_{\G_A} ={\phi}^{A_2}_i\vert_{\G_A} \,, \qquad {\phi'}^{A_2}_i\vert_{\G_A} ={\phi'}^{A_1}_i\vert_{\G_A} \,, \qquad i=1, \dots, \nev\,,
\ee
and at the cut between $A_a$ and $B$, denoted $\G_a$, they are 
\be
\phi^{A_a}_i\vert_{\G_a} ={\phi}^{B}_i\vert_{\G_a} \,, \qquad {\phi'}^{A_a}_i\vert_{\G_a} ={\phi'}^{B}_i\vert_{\G_a} \,, \qquad a=1,2\,, \quad i=1, \dots, \nev\,. 
\ee
When we combine the gluing and continuity conditions, we see that the fields must satisfy
\begin{equation}
\label{B-no-winding}
\B:\qquad
\begin{array}{lc}
\begin{aligned}[l]
       \phi^{A_a}_i\vert_{\Gamma_a}=\phi^{B}_j\vert_{\Gamma_a} = \chi_a\,,\\
    %   \phi^{A_2}_i\vert_{\Gamma_2}=\phi^{B}_j\vert_{\Gamma_2}\equiv \chi_2\\
\end{aligned}\;& \qquad\; a=1, 2\,, \quad i,j=1,\ldots,n_e\,,\vspace{3pt}\\
\left.\begin{aligned}
       & \phi^{A_1}_{k}\vert_{\Gamma_A}=\phi^{A_2}_{\ell}\vert_{\Gamma_A} = \chi_A^{o}\\
      & \phi^{A_1}_{n_e/2+k}\vert_{\Gamma_A}=\phi^{A_2}_{n_e/2+\ell}\vert_{\Gamma_A} = \chi_A^e\\
\end{aligned}\;\right\rbrace &   \hspace*{-18pt}  k,l=1,\ldots,n_e/2\,,
\end{array}
\end{equation}
as depicted in Fig.~\ref{fig:gluing-adjacent}. 
Notice that we have relabelled the $\nev$ independent fields %with support on $A$ \CB{only on A?} 
in order to have the odd fields ranging from $1$ to $n_e/2$ and the even fields from $n_e/2+1$ to $n_e$.
The functions $\chi_a, \chi_A^{o}, \chi_A^{e}$ are arbitrary and only defined at the corresponding entangling cuts, essentially by the above conditions. 

The main difference compared to the case of disjoint submanifolds now becomes apparent:
We have two independent sets of $n_e/2$ boundary conditions at the entangling cut between $A_1$ and $A_2$, while at the other cuts $\G_1$ and $\G_2$ we still have a single set of $\nev$ conditions. This means that, contrary to the disjoint case, the adjacent geometry is sensitive to the partial transposition.

Using the boundary conditions $\mathcal B$ given in \eqref{B-no-winding}, we can now directly write 
\begin{equation}
\label{tr-rho-A-partial-transpose-A2}
   \tr\big(\rho_{A}^{T_{2}}\big)^{n_e} =\frac{1}{Z_{A\cup B}^{n_e}}\int_{\mathcal{B}}\prod_{i=1}^{n_e}\D\phi_i^{A_1}\, e^{-S[\phi_i^{A_1}]}\int_{\mathcal{B}}\prod_{i=1}^{n_e}\D\phi_i^{A_2}\, e^{-S[\phi_i^{A_2}]}\int_{\mathcal{B}}\prod_{i=1}^{n_e}\D\phi_i^{B}\, e^{-S[\phi_i^{B}]}\,.
\end{equation}
For a pure state, the path integrals factorise in a straightforward way at this point. 
The situation here is a little more complicated since the entangling cuts carry different numbers of degrees of freedom -- one at $\Gamma_{1,2}$ and two at $\Gamma_A$. 
However, this difficulty may be circumvented by rotating the fields as described originally in \cite{Fradkin:2006mb,Zaletel2011,Zhou:2016ykv}.
Let us first define a unitary rotation matrix $U_n$ \cite{Zhou:2016ykv} as follows
\be\label{def-matrix-U}
U_n= 
\begin{bmatrix}
{1\over \sqrt 2} ~&~ -{1\over \sqrt{2}}~ &~0 & ~&\dots\\
{1\over \sqrt 6} ~& ~{1\over \sqrt{6}} ~ & ~ - {2\over \sqrt{6}} & ~~0& \dots \\
\vdots 
\\
{1\over \sqrt{n(n-1)}} ~& ~{1\over \sqrt{n(n-1)}} & \dots~ & ~\dots & -\sqrt{1-{1\over n}}\\
{1\over \sqrt{n}} ~& ~{1\over \sqrt{n}} & \dots~ & ~\dots & {1\over \sqrt{n}}\\
\end{bmatrix}.
\ee
It is chosen such that the first $n-1$ rotated fields vanish on the entanglement cuts.
Two rotations are then performed independently on the first and on the last $n_e/2$ fields with the help of the block diagonal matrix $\tilde{U}_{n_e}= U_{n_e/2}\oplus U_{n_e/2}$. In vector notation this rotation reads $\tildephi = \tilde{U}_{n_e}\phi$ and results in the boundary conditions   
\begin{equation}
\label{Btilde-no-winding}
\tilde{\B}:\qquad
\begin{array}{lc}
\begin{aligned}[l]
\tildephi^{A_a}_i\vert_{\Gamma_a}=\tildephi^{B}_j\vert_{\Gamma_a}= \sqrt{\frac{n_e}{2}} \chi_a\,,\;\\
%\tildephi^{A_2}_i\vert_{\Gamma_2}=\tildephi^{B}_j\vert_{\Gamma_2}= \sqrt{\frac{n_e}{2}}\chi_2\;\\
\end{aligned} &\quad a=1,2 \,,  \quad i,j=n_e/2, n_e\vspace{2pt}\\
\displaystyle \tildephi^{A_1}_{n_e/2}\vert_{\Gamma_A}=\tildephi^{A_2}_{n_e/2}\vert_{\Gamma_A}=\sqrt{\frac{n_e}{2}} \chi_A^{o}\,,\vspace{2pt}\\
\displaystyle \tildephi^{A_1}_{n_e}\vert_{\Gamma_A}=\tildephi^{A_2}_{n_e}\vert_{\Gamma_A}=\sqrt{\frac{n_e}{2}} \chi_A^e\,, & 
\end{array}
\end{equation}
with the remaining fields vanishing at all cuts. We then perform an additional $U_2$ rotation on the fields $\tildephi_{\nev/2}, \tildephi_{\nev}$ as to obtain
\be
\label{Bbar-no-winding}
\bar{\B}:\qquad
\begin{aligned}
&\barphi^{A_a}_{n_e}\vert_{\Gamma_a}=\barphi^{B}_{n_e}\vert_{\Gamma_a} = \sqrt{n_e}\hp\chi_a\,, \qquad a=1,2 \,,\vspace{3pt}\\
%&\barphi^{A_2}_{n_e}\vert_{\Gamma_2}=\barphi^{B}_{n_e}\vert_{\Gamma_2}\equiv \sqrt{n_e}\chi_2\,,\\
&\barphi^{A_1}_{n_e}\vert_{\Gamma_A}=\barphi^{A_2}_{n_e}\vert_{\Gamma_A} = \sqrt{\frac{n_e}{2}}\chi_+\,,\\
&\barphi^{A_1}_{n_e/2}\vert_{\Gamma_A}=\barphi^{A_2}_{n_e/2}\vert_{\Gamma_A}= \sqrt{\frac{n_e}{2}} \chi_-\,,
\end{aligned}
\ee
where $$\chi_\pm :=\frac{1}{\sqrt{2}}(\chi_A^o\pm\chi_A^e)$$ are again arbitrary and independent functions, and the  $\nev-2$ remaining fields have Dirichlet boundary conditions at all entangling cuts. 

Each of these $\nev-2$ fields thus produces three Dirichlet partition functions in \eqref{tr-rho-A-partial-transpose-A2}: on $A_1$, $A_2$ and $B$. 
Further inspection of the boundary conditions \eqref{Bbar-no-winding} reveals that the $n_e$--th field is free on the whole 
manifold $A\cup B$. It is not constrained to vanish at any cut, and the cut functions $\chi$ are arbitrary, which allows us to write
\footnote{
There are additional non-universal factors $\propto n_e^{-\text{L}(\Gamma_i)/2\eps}$, where $\text{L}(\Gamma_i)$ is the length of $\Gamma_i$ and $\eps$ a UV-cutoff, to the partition functions \eqref{ZABadjNC} and \eqref{ZAadjNC}, arising from the Jacobian that results of the successive rotations applied to the fields, see \cite{Zaletel2011,Zhou:2016ykv}. They only contribute to the area law and can thus be ignored.}
\be
\lb{ZABadjNC}
&& \hspace{-5pt} Z_{A\cup B}\, =\nn
\\ \nn
&&\int_{\bar{\B}}\hspace{-2pt}\D\barphi_{n_e}^{A_1}\, e^{-S[\barphi_{n_e}^{A_1}]}\hspace{-3pt}\int_{\bar{\B}}\hspace{-2pt}\D\barphi_{n_e}^{A_2}\, e^{-S[\barphi^{A_2}_{n_e}]}\hspace{-3pt}\int_{\bar{\B}}\hspace{-2pt}\hspace{-2pt}\D\barphi_{n_e}^{B}\, e^{-S[\barphi_{n_e}^{B}]}
\hspace{-3pt}\int \hspace{-2pt}[\D\barphi_{n_e}\vert_{\G_A}] \, e^{-S[\barphi_{n_e}]} \prod_{a}\int \hspace{-2pt}[\D\barphi_{n_e}\vert_{\G_a}] \, e^{-S[\barphi_{n_e}]}\,,\\
\ee
where the last integrals indicate the sum over all possible values of the field at the entangling cuts~\cite{Zhou:2016ykv,Angel-Ramelli:2019nji}. 
The $n_e/2$--th field is only free on $A=A_1\cup A_2$ (it is not subject to Dirichlet boundary condition only at $\Gamma_A$) once we sum over the degrees of freedom along the cut $\G_A$, such that 
\be
\lb{ZAadjNC}
&& \hspace{-17pt} Z_{A_1\cup A_2}Z_B\, = \nn
\\ \nn
&& \int_{\bar{\B}}\D\barphi_{n_e/2}^{A_1}\, e^{-S[\barphi_{n_e/2}^{A_1}]}\int_{\bar{\B}}\D\barphi_{n_e/2}^{A_2}\, e^{-S[\barphi^{A_2}_{n_e/2}]}\int_{\bar{\B}}\D\barphi_{n_e/2}^{B}\, e^{-S[\barphi_{n_e/2}^{B}]} \int [\D\barphi_{n_e/2}\vert_{\G_A}] \, e^{-S[\barphi_{n_e/2}]}\,. \\
\ee
Here the partition function over $B$ is calculated with Dirichlet boundary conditions and the one over $A$ with boundary conditions dictated by the geometry in question, as we will discuss in detail below. 
Hence,  we can finally rewrite \eqref{tr-rho-A-partial-transpose-A2} as 
\begin{equation}
\label{tr-rho-A-partial-transpose-A2-3}
\tr\big(\rho_{A}^{T_{2}}\big)^{n_e} =\frac{(Z_{A_1}Z_{A_2}Z_B)^{n_e-2}Z_{A\cup B}Z_{A_1\cup A_2}Z_{B}}{Z_{A\cup B}^{n_e}}\,, 
\end{equation}
where the partition functions over $A_1$ and $A_2$ separately are also computed assuming Dirichlet boundary conditions. 
The resulting logarithmic negativity \eqref{LNreplica} is given by
\begin{equation}
\label{curly-E-gen-replica-nowinding}
\E=-\log(\frac{Z_{A_1}Z_{A_2}}{Z_{A_1\cup A_2}})\,. 
\end{equation}
As before, we expect this formal expression to be valid for $(d+1)$-dimensional Lifshitz theories with even exponent $z$ on flat space and with some caveats on curved manifolds, such as $z\le d$ for the sphere. 

Notice that in analogy to the entanglement entropy~\cite{Fradkin:2006mb,Zaletel2011,Zhou:2016ykv,Angel-Ramelli:2019nji}, the logarithmic negativity turns out to be a difference of free energies between the two subsystems involved and their union, confirming our expectation from Section \ref{sec:hint-at-a-general-formula}.

%%%%%%%%%%%%%%%%%%%%%%%%%%%%%%%%%%%%%%%%
%%%%%%%%%%%%%%%%%%%%%%%%%%%%%%%%%%%%%%%%
\subsection{Adjacent submanifolds with winding}
\label{sec:adjacent-manifolds-winding}

The basic procedure that we used in the previous section carries through to compact fields, that is fields with $\phi\sim\phi+2\pi R_c$. However, as a consequence of the compact nature of $\phi$, the boundary conditions \eqref{B-no-winding} need only be satisfied modulo $2\pi R_c$. The periodic identification is taken into account in the standard way~\cite{Ginsparg1988,DiFrancesco1997}, by writing each replicated field as a sum of a classical field and a fluctuation, $\phi = \phi^\cl+\varphi$. 
The classical field obeys the equations of motion and 
takes the value of the total field at the entangling cuts, including any winding contribution, while the fluctuation satisfies Dirichlet boundary conditions at all the cuts. 
This definition ensures that the action factorises as $S[\phi]=S[\phi^\cl]+S[\varphi]$, and we can rewrite our path integrals as
\begin{equation}
\int\D\phi_i\ e^{-S[\phi_i]}=\int\D\varphi_i\,e^{-S[\varphi_i]}\sum_{\phi_i^\cl} e^{-S[\phi_i^\cl]}
\end{equation}
 for each field $i=1,\cdots,n_e$. 
The classical fields satisfy the boundary conditions
\be
\phi_i^\cl \vert_{\Gamma_a}=\chi_a+2\pi R_c\, \omega^a_i,\qquad \omega^a_i\in\Z\,,
\ee
instead of \eqref{B-no-winding}, where $a=1, 2, A$ labels the cut $\Gamma_a$. 
The field $\phi_i^\cl$ is defined on the complete manifold. It is found by solving the equations of motion on each submanifold and stitching the resulting fields together across the cuts, subject to the above boundary conditions.
Furthermore, depending on the global symmetries of the geometry, some of the winding modes $\omega_i^a$ may be redundant. This means that one needs to specify a geometry from the start, carefully identify the non-redundant winding modes, and only sum over these when performing the path integral manipulations of the last section. In the end, this procedure leads to a logarithmic negativity of the form
\begin{equation}
\label{curly-gen-replica-winding}
\E=-\log(\frac{Z_{A_1}Z_{A_2}}{Z_{A_1\cup A_2}})+\log W_{\hspace{-1pt}\E}(1)\,,
\end{equation}
where $W_{\hspace{-1pt}\E}(n)$ is the contribution from the winding sector encoding the topological information that resides in the classical fields. We note that $W_\E$ is heavily dependent on the geometry as is illustrated below via explicit examples. 

%%%%%%%%%%%%%%%%%%%%%%%%%%%%%
\subsubsection{Spherical geometry}
\label{sec:sphere-logneg}

Let us consider the spherical configuration on the left in Fig.~\ref{fig:realizations-of-adjacent-geometry}.  
There are only two cuts, $\Gamma_A$ and $\Gamma_2$, but the previous formulae carry over if we simply ignore the trivial 
cut $\Gamma_1$. A priori, we have $2n_e$ winding numbers $\omega_i^a$: one for each replica (labeled by $i=1,\ldots,n_e$) at each cut (labeled by $a=\{2,A\}$).  
We also have three arbitrary functions, $\chi_2$ defined along $\G_2$ and $\chi_e, \chi_o$ along $\G_A$, which can be 
redefined so as to absorb one winding mode each. 
In what follows, we choose to eliminate the $\nev$--th mode at the cut $\G_2$, and the $\nev/2$--th and $\nev$--th modes at $\G_A$. 
In addition, the sphere admits a global shift symmetry, $S[\phi]=S[\phi+\text{const.}]$, 
which we can use to get rid of all the remaining winding modes at $\G_2$. 
Since the global shift affects all cuts uniformly, the winding numbers at $\Gamma_A$ get shifted to $\omega_i^A - \omega_i^2$, but we can, without loss of generality, relabel them as $\omega_i^A$ to avoid cluttering the notation.
We thus end up with only $n_e-2$ of the original $2n_e$ winding modes. The boundary conditions for the classical fields turn into 
\begin{equation}
\label{B-sphere}
\B:\qquad
\begin{array}{lc}
\hspace{1pt}\phi^\cl_i\vert_{\Gamma_2}= \chi_2\,,&   \hspace*{-16pt} i=1,\ldots,\nev\,, \\
\left.\begin{aligned}
   &\phi^\cl_{k}\vert_{\Gamma_A}= \chi_A^o+2\pi R_c\hp\omega^A_{k}\,,\\
   &\phi^\cl_{n_e/2+k}\vert_{\Gamma_A}= \chi_A^e+2\pi R_c\hp \omega^A_{n_e/2+k}\,,\\
\end{aligned}\right\rbrace & \quad k=1,\ldots,n_e/2-1\,,\\
\hspace{1pt}\phi^\cl_{n_e/2}\vert_{\Gamma_A}=\chi_A^o\,,& \\
\hspace{1pt}\phi^\cl_{n_e}\vert_{\Gamma_A}=\chi_A^e\,,&
\end{array}
\end{equation}
while the fluctuations have Dirichlet boundary conditions at all cuts, 
$$\varphi_i\vert_{\Gamma_a}=0\,, \qquad i=1,\cdots,n_e\,.$$
We proceed exactly as in Section~\ref{sec:adjacent-no-winding} and perform first a rotation $\tilde{U}_{n_e}$ of all the fields, followed by an additional $U_2$ rotation of the $n_e/2$--th and $n_e$--th fields. We obtain boundary conditions analogous to \eqref{Bbar-no-winding}, except that now there is also a winding sector contribution at $\Gamma_A$, 
\begin{equation}
\label{B-sphere2}
\bar{\B}:\qquad
\begin{array}{lc}
\hspace{1pt}\displaystyle\barphi^\cl_{n_e}\vert_{\Gamma_2}= \sqrt{n_e}\hp\chi_2\,,&\vspace{5pt} \\
\left.\begin{aligned}
&\displaystyle\barphi^\cl_j\vert_{\Gamma_A}= 2\pi R_c ({U}_{n_e/2})_{jk}\hp\omega_k^A\,,\\
&\displaystyle\barphi^\cl_{\nev/2+j}\vert_{\Gamma_A}= 2\pi R_c ({U}_{n_e/2})_{jk}\hp\omega_{\nev/2+k}^A\,,\\\end{aligned}\right\rbrace & \hspace{-1.5cm} j,k=1,\ldots,n_e/2-1\,,\\
\hspace{1pt}\displaystyle\barphi_{\nev}^\cl\vert_{\Gamma_A}=\sqrt{\frac{n_e}{2}}\chi_++ \frac{2\pi R_c}{\sqrt{\nev}}\sum_{i=1}^{n_e/2-1}\hspace{-3pt}\left(\omega^A_{i}+\omega^A_{i+\nev/2}\right),&\vspace{2pt}\\
\hspace{1pt}\displaystyle\barphi_{\nev/2}^\cl\vert_{\Gamma_A}=\sqrt{\frac{n_e}{2}}\chi_-+ \frac{2\pi R_c}{\sqrt{\nev}}\sum_{i=1}^{\nev/2-1}\hspace{-3pt} \left(\omega^A_{i}-\omega^A_{i+\nev/2}\right),&
\end{array}
\end{equation}
where $\chi_\pm=\frac{1}{\sqrt{2}}(\chi_A^o\pm\chi_A^e)$ and with the first $n_e-1$ classical modes vanishing at $\Gamma_2$.

Just as in the non-compact case (see equations \eqref{ZABadjNC} and \eqref{ZAadjNC})  we can then use the $n_e$--th mode to reconstruct a full partition function on the sphere, while the $n_e/2$--th serves to reconstruct a partition function on $A=A_1\cup A_2$ with Dirichlet boundary conditions at the boundary $\G_2$. 
The remaining $n_e-2$ fluctuating fields lead to Dirichlet partition functions on each submanifold, $A_1$, $A_2$ and $B$, and the classical modes yield pure winding sums.
We thus arrive at the expression
\begin{equation}
\label{tr-rho-A-partial-transpose-A2-sphere-2}
\tr\big(\rho_{A}^{T_{2}}\big)^{n_e} = \sqrt{\nev}\, \frac{(Z_{A_1}Z_{A_2}Z_B)^{n_e-2}Z_{A\cup B}Z_{A_1\cup A_2}Z_{B}}{Z_{A\cup B}^{n_e}}\,\WE(n_e)\,,
\end{equation}
with the winding sector given by
\begin{equation}
\label{newequationname}
\WE(n)=\sum_{\omega^A\in\Z^{n_e-2}}e^{-\sum_{i\neq n_e/2,n_e}S[\barphi^\cl_i]}\,.
\end{equation}
$\WE$ is constructed in terms of the $n_e-2$ classical fields satisfying the conditions \eqref{B-sphere2}, which can be summarised as follows:
\be
&&\hspace{1pt}\barphi^\cl_i\vert_{\Gamma_2}\,=0 \,, \qquad\;\;   i=1, \dots, \nev-1\, \;\;\;\text{with} \quad i\neq\nev/2\,,\nn\vspace{2pt} \\
&&\left.\begin{aligned}
&\barphi^\cl_j\vert_{\Gamma_A}= 2\pi R_c (M_{\nev/2-1})_{jk}\hp\omega_k^A\,, \\
&\barphi^\cl_{\nev/2+j}\vert_{\Gamma_A}= 2\pi R_c (M_{n_e/2-1})_{jk}\hp\omega_{\nev/2+k}^A\,,
\end{aligned}\right\rbrace \quad j,k=1,\ldots,\nev/2-1\,,
\lb{B-sphere-winding-V}
\ee
where we have introduced a matrix $M_{m-1}$ obtained by deleting the $m$--th row and column of $U_m$ in \eqref{def-matrix-U}, as was done in \cite{Zhou:2016ykv}.
The factor $\sqrt{\nev}$ in \eqref{tr-rho-A-partial-transpose-A2-sphere-2} is essentially due to the global shift symmetry forcing all the classical fields at the entangling cut $\G_2$ to be the same~\cite{Zhou:2016ykv, Angel-Ramelli:2019nji}. 
Consequently, the $\nev$--th classical field gets its compactification radius amplified by $\sqrt{\nev}$, which in turn needs to be compensated for in the partition function. 
The factor of $\sqrt{\nev}$ does not contribute to the logarithmic negativity but is crucial for getting the correct entanglement entropy on hemispheres and tori~\cite{Zhou:2016ykv, Angel-Ramelli:2019nji}.  

Thanks to the factorisation of the boundary conditions \eqref{B-sphere-winding-V}, the winding sector contribution \eqref{newequationname} can be expressed as 
\be
\label{winding-2sphere-imp}
\WE(\nev)=\left(\sum_{\omega^A\in\Z^{n_e/2-1}}e^{-\sum_{j=1}^{n_e/2-1}S[\barphi^\cl_j]}\right)^2\, = W(n_e/2)^2 \,,
\ee
where each set of $\nev/2$--fields produces the winding sector $W(n_e/2)$ appearing in the entanglement entropy calculated in \cite{Zhou:2016ykv}.

Classical solutions satisfying the boundary conditions  \eqref{B-sphere-winding-V}  can be obtained via a conformal transformation which projects spherical caps to annuli~\cite{Zhou:2016ykv}.
With the annulus radial coordinate $\eta$ given by 
$
\eta=\tan{\hspace{-1pt}(\theta/2)}
$,
where $\theta$ is the polar angle in spherical coordinates, the classical solution reads
\be
\phi^\cl(\eta)={\phi^\cl_{|\G_A}\over \log{{\eta_A\over \eta_2}}} \log{{\eta\over \eta_2}}\,,
\ee
where  $\eta_{A (2)}= \tan{\hspace{-1pt}(\theta_{A(2)}/2)}$ correspond to the positions of the entangling cuts $\Gamma_{A(2)}$ on the sphere%
\footnote{Only classical fields with support in the region $A_2$ (see Fig.~\ref{fig:realizations-of-adjacent-geometry}) are non-zero, and thus contribute to the winding sector~\cite{Zhou:2016ykv}.}.
The analytic continuation of $W(n)$ obtained in \cite{Zhou:2016ykv} is given by
\be\label{w-annulus}
W(n)= \sqrt n \hp c^{-{{n-1}\over2}} \int_{-\infty}^{\infty}{dk\over \sqrt \pi} e^{-k^2} \left[ \sum_{\omega\in\Z} \exp\left(-{\pi\over c} \omega^2 -2 i \sqrt{{\pi\over c}}k \omega\right)\hspace{-1pt}\right]^{n-1},
\ee
where the constant $c$ takes the value 
$$c={8 \pi^2 R^2_c g\over \log{\hspace{-1pt}(\eta_2/\eta_A)}}.$$
Hence, from \eqref{tr-rho-A-partial-transpose-A2-sphere-2} and \eqref{winding-2sphere-imp} we obtain for the logarithmic negativity  
\be\label{log-neg-sphere-implicit}
\E = - \log(\frac{Z_{A_1}Z_{A_2}}{Z_{A_1\cup A_2}})+2 \log W(1/2)\,. 
\ee
The partition functions in \eqref{log-neg-sphere-implicit} can be computed via functional determinants and regularised by means of zeta-function techniques. 
For the regularised functional determinants we use the results of~\cite{Weisberger:1987kh,Dowker:1993hq,Zhou:2016ykv}, reported in Appendix \ref{app:det-sphere} below,  where $A_1$ and $A_1\cup A_2$ are spherical caps with Dirichlet conditions at the boundary, and $A_2$ is the ``belt'' region between two spherical caps ($A_1$ and $B$) with Dirichlet conditions at both boundaries. 
This yields 
\be
\label{det-2-sphere-exp}
 - \log(\frac{Z_{A_1}Z_{A_2}}{Z_{A_1\cup A_2}}) = {1\over 2} \log \log{\eta_2\over \eta_A} -\half \log \pi +\half \sum_{m>0} \log\hspace{-1pt}\left(1-\Big({\eta_A\over \eta_2}\Big)^{2m}\right)^2,
\ee
while for the winding sector we have
\be
&&\hspace{-5pt}2 \log W(1/2)=\nn\\
&&\quad\log {\sqrt{2 \pi^2 g} R_c}-\half \log \log  {\eta_2\over \eta_A}+2 \log \int_{-\infty}^\infty {dk\over \sqrt \pi} e^{-k^2}{\left[ \sum_{\omega\in\Z} \exp\hspace{-1pt}\left(-{\pi\over c} \omega^2 -2 i \sqrt{{\pi\over c}}k \omega\right)\right]^{-1/2}}.\quad\nonumber\\
\ee
Putting everything together, our explicit analytic expression for the logarithmic negativity is given by 
 \be
 \label{log-neg-final-2sphere-exp}
 \E &=&\log {\sqrt{2 \pi g} R_c} +\half \sum_{m>0} \log\hspace{-1pt}\left(1-\Big({\eta_A\over \eta_2}\Big)^{2m}\right)^2\nn
 \\
 &&+2 \log \int_{-\infty}^\infty {dk\over \sqrt \pi} e^{-k^2}{\left[ \sum_{\omega\in\Z} \exp\hspace{-1pt}\left(-{\pi\over c} \omega^2 -2 i \sqrt{{\pi\over c}}k \omega\right)\right]^{-1/2}}.
 \ee
Let us consider the pure state regime, where $B\to \emptyset$. This corresponds to the limit $\eta_2 \gg \eta_A$, that is $c\ll 1$, and we obtain 
\be\label{pure-state-limit-LN-sphere}
\E =\log {\sqrt{2 \pi g} R_c} -2 \left({\eta_A\over \eta_2}\right)^{\frac{1}{4 \pi  g R_c^2}}- \left({\eta_A\over \eta_2}\right)^2+\dots\,,
\ee
since
\be
2  \log W(1/2) 
&=& \log{\sqrt{{2 \pi^2 g}}R_c}-\half \log \log{{\eta_2\over \eta_A}}- 2 \left({\eta_A\over \eta_2}\right)^{\frac{1}{4 \pi  g R_c^2}}+\cdots\,.
\ee
We expect to recover the $1/2$--th R\'enyi entropy in that case, as discussed in Section \ref{sec:logarithmic-negativity-of-pure-states}. 
Indeed, from \cite{Zhou:2016ykv, Angel-Ramelli:2019nji} one has for pure states
\be
\nn
S_{A_1}^{(1/2)}= \log\sqrt{2 \pi g} R_c\,,
\ee
which is in agreement with \eqref{pure-state-limit-LN-sphere}. 
We note, however, a curious difference in the dependence on the compactification radius between the R\'enyi entropy and the 
logarithmic negativity. On the one hand, in the pure state, the dependence of the R\'enyi entropy and entanglement entropy on the compactification radius $R_c$ is due to a zero mode of the partition function on the sphere~\cite{Zhou:2016ykv, Angel-Ramelli:2019nji}. 
On the other hand, the partition functions in \eqref{log-neg-sphere-implicit} and \eqref{det-2-sphere-exp} have no zero modes and
therefore the same dependence in the logarithmic negativity $\E$ must arise from the winding sector.

%%%%%%%%%%%%%%%%%%%%%%%%%%%%%%%%%%%%%%%%
%%%%%%%%%%%%%%%%%%%%%%%%%%%%%%%%%%%%%%%%

\subsubsection{Toroidal geometry}
\label{sec:torus-logneg}

The toroidal geometry is shown in the right image of Fig.\,\ref{fig:realizations-of-adjacent-geometry}. We consider the torus to have area $L_1\cross L_2$ and to be cut along the $L_1$ direction. We denote the lengths of $A_1$, $A_2$, and $B$ by $\ell_1$, $\ell_2$, and $\ell_B$ respectively.
As before we partially trace over $B$ first, and then we partially transpose over $A_2$. 
The  first step is to understand how the boundary conditions \eqref{B-no-winding} are modified by the presence of the winding modes. 
Each replicated field is split in classical and fluctuating fields, where the latter obey Dirichlet boundary conditions at the cuts. Hence, the only remaining task is understanding the boundary conditions obeyed by the classical fields $\phi^\cl$. 
In a torus the minimal choice we can do requires three cuts and thus a priori $3 \nev$ winding numbers. 
From the discussion in Section \ref{sec:adjacent-no-winding} we know that there are four cut functions: $\chi_1$ ($\chi_2$) for the cut $\G_1$ ($\G_2$) between $A_1$ ($A_2$) and $B$, and $\chi^{e/o}$ defined at the entangling cut $\G_A$ between $A_1$ and $A_2$.
As in the spherical case, we redefine the $\chi$ functions at $\Gamma_1$, $\Gamma_2$ and $\Gamma_A$ to absorb the $\nev$--th winding mode of each of the cuts, as well as the $\nev/2$--th winding mode in $\Gamma_A$.
The torus admits a global shift symmetry 
$\phi_i\to \phi_i+\text{const.}$, % as highlighted in boundary conditions above, 
which we use it to get rid of all the $\nev-1$ winding modes at the cut $\G_2$ (the choice between $\G_1$ and $\G_2$ is completely equivalent). 
Relabelling the difference of winding modes, we end up with the following boundary conditions
\begin{subequations}
\label{B-torus-new-notation-v1}
\begin{align}
 &\phi^\cl_i\vert_{\Gamma_2} = \chi_2\,, \qquad \hspace{2.51cm} i=1, \dots, \nev \,, \\ 
 & \phi^\cl_i\vert_{\Gamma_1} = 
\begin{cases}
\chi_1 + 2 \pi R_c\hp  \omega_i^{1}\,, \qquad &\hp i=1, \dots, {\nev} -1\,,\\
\chi_1\,,\qquad &\hp i=\nev\,, 
 \end{cases}
\\ 
& \phi^\cl_i\vert_{\Gamma_A} = 
\begin{cases}
\chi_o + 2 \pi R_c\hp  \omega_i^{o}\,, \qquad & i=1, \dots, {\nev/2} -1\,,\\
\chi_o\qquad & i=\nev/2\,, 
\end{cases}\\
\nn
& \phi^\cl_{i+\nev/2}\vert_{\Gamma_A} = \begin{cases}
\chi_e + 2 \pi R_c\hp \omega_i^{e}\,, \qquad & i=1, \dots, {\nev/2}-1\,,\\
\chi_e\,, \qquad  &i=\nev/2\,,
 \end{cases}
\end{align}
\end{subequations}
with $\omega^a_i\in\Z$. From the above conditions, it is clear that we have $2\nev-3$ independent modes, which are not symmetrically distributed among the cuts: the two cuts $\G_1, \G_A$ carry a different number of degrees of freedom, and thus of winding modes, which ultimately gives rise to a rather involved expression for the winding sector. 
As done previously, we rotate the classical fields with a $\tilde U_{\nev}$ rotation that acts separately on the first and second sets of $\nev/2$ fields, see equation \eqref{Btilde-no-winding}, and obtain 
\begin{subequations}
\label{B-torus-new-notation-v2}
\begingroup
\allowdisplaybreaks
\begin{align}
 \tilde\phi^\cl_i\vert_{\Gamma_2} &=  
 \begin{cases}
 \sqrt{{\nev\over 2}}\chi_2 \qquad 
& i= \nev/2\,,  \nev \,, \\ 
0\,, \qquad & i=1, \dots, \nev-1, \quad \text{with} \quad  i\neq\nev/2\,, 
\end{cases}
\\[10 pt]
\label{B-torus-new-notation-v2-2}
 \tilde \phi^\cl_i\vert_{\Gamma_1} &= 
\begin{cases}
2 \pi R_c (M_{\nev/2-1})_{ij} \omega^1_j -2 \pi R_c  \sqrt{1-{2\over \nev}} \delta_{i, \nev/2-1} \omega^1_{\nev/2}\,, \quad &i, j=1, \dots, \nev/2-1\,, \\
 \sqrt{{\nev\over 2}} \chi_1 + {2 \pi \, R_c\over \sqrt{\nev/2}}\, \sum_{k=1}^{\nev/2}  \omega_k^{1}\,, \quad & i= {\nev/2}\,,
 \end{cases}
 \\ \nn
  \tilde \phi^\cl_{i+\nev/2}\vert_{\Gamma_1} &= 
  \begin{cases}
 2 \pi R_c (M_{\nev/2-1})_{ij} \omega^1_{j+\nev/2}\,, \hspace{1.45cm} &  i, j=1, \dots, \nev/2-1\,, \\
 \sqrt{{\nev\over 2}}   \chi_1 +\sum_{k=1}^{\nev/2-1} \omega^1_{k+\nev/2} \qquad & i=\nev/2 \,, 
 \end{cases}
\\[10 pt]
\label{B-torus-new-notation-v2-3}
 \tilde\phi^\cl_i\vert_{\Gamma_A} &= 
\begin{cases}
2 \pi R_c (M_{\nev/2-1})_{ij} \omega^o_j  \qquad & i, j=1, \dots, {\nev/2} -1\,,\\
 \sqrt{{\nev\over 2}} \chi_o + {2 \pi \, R_c\over \sqrt{\nev/2}}\, \sum_{k=1}^{\nev/2-1}   \omega_k^{o}\,, \qquad & i= {\nev/2}\,,
 \end{cases}
 \\\nn
  \tilde\phi^\cl_{i+\nev/2}\vert_{\Gamma_A} &= 
 \begin{cases}
 2 \pi R_c (M_{\nev/2-1})_{ij} \omega^e_j  \qquad & i, j=1, \dots, {\nev/2} -1\,,\\
 \sqrt{{\nev\over 2}}  \chi_e + {2 \pi \, R_c\over \sqrt{\nev/2}}\, \sum_{k=1}^{\nev/2-1}   \omega_k^{e}\,, \qquad & i= {\nev/2}\,.\\
 \end{cases}
\end{align}
\endgroup
\end{subequations}
The ``unusual'' first line in \eqref{B-torus-new-notation-v2-2} is simply due to the explicit form of the matrix $U_{\nev/2}$. 
The winding mode $\omega_{\nev/2}^1$ is responsible for the coupling between the various frequencies, as will become clear later. 
In order to simplify the notation, we define the following  vectors 
\begin{align}
 &\mu^1 :=  (\omega^1_1, \dots, \omega^1_{\nev/2-1}), & &\upsilon^1 :=  (\omega^1_{\nev/2+1}, \dots, \omega^1_{\nev-1}), \nonumber\\
 &\mu^A :=  (\omega^o_1, \dots, \omega^o_{\nev/2-1}), & &\upsilon^A :=  (\omega^e_{\nev/2+1}, \dots,\omega^e_{\nev-1}),\\
 &\gamma  := \omega^1_{\nev/2}, & &I:= (1, \dots, 1)\in \Z^{\nev/2 -1}\,. \nonumber
\end{align}
Performing the last $U_2$ rotation on the fields labelled by $\nev$ and $\nev/2$, we obtain 
\begin{subequations}
\label{B-torus-new-notation-v3}
\begingroup
\allowdisplaybreaks
\begin{align}
 \bar\phi^\cl_i\vert_{\Gamma_2}  &=  
 \begin{cases}
 0 \,, \qquad  &i=1, \dots, \nev-1\,,\\
 \sqrt{{\nev}}\chi_2 \,,\qquad 
& i= \nev \,, 
 \end{cases}
\\[15pt]
\label{B-torus-new-notation-v3-2}
 \bar \phi^\cl_i\vert_{\Gamma_1} &= 
\begin{cases}
2 \pi R_c (M_{\nev/2-1})_{ij} \mu^1_j - 2 \pi R_c \sqrt{1-{2\over \nev}} \delta_{i, \nev/2-1} \gamma\,, \qquad &i, j=1, \dots, \nev/2-1\,, \\
  {2 \pi \, R_c\over \sqrt{\nev}}\, \left(I \cdot \mu^1-I\cdot \upsilon^1 +\gamma \right)\,, \qquad & i= {\nev/2}\,,
  %\sum_{i=1}^{\nev/2-1} \left(\mu^1_i -  \upsilon_{i}^{1}\right) +   {2 \pi \, R_c\over \sqrt{\nev}}\,\gamma\,, \qquad & i= {\nev/2}\,,\\
  \end{cases}
  \\ \nn
   \bar \phi^\cl_{i+\nev/2}\vert_{\Gamma_1} &= 
\begin{cases}
 2 \pi R_c (M_{\nev/2-1})_{ij} \upsilon^1_{j}\,, \qquad &  i, j=1, \dots, \nev/2-1\,, \\
 \sqrt{{\nev}} \chi_1+   {2 \pi \, R_c\over \sqrt{\nev}}\, I\cdot  \upsilon^{1}\,, \qquad & i=\nev/2\, \,, 
 % {2 \pi \, R_c\over \sqrt{\nev}}\, \sum_{i=1}^{\nev-1} \upsilon^{1}_i\,, \qquad & i=\nev\, \,, 
 \end{cases}
\\[15 pt] 
\label{B-torus-new-notation-v2-3}
 \bar\phi^\cl_i\vert_{\Gamma_A} &= 
\begin{cases}
2 \pi R_c (M_{\nev/2-1})_{ij} \mu^A_j  \,, \qquad & i, j=1, \dots, {\nev/2} -1\,,\\
 \sqrt{{\nev\over 2}}  \chi_- + {2 \pi \, R_c\over \sqrt{\nev}}\, \left( I\cdot \mu^A - I\cdot \upsilon^A\right)\,, \qquad & i= {\nev/2}\,,
 % {2 \pi \, R_c\over \sqrt{\nev}}\, \sum_{i=1}^{\nev/2-1}  \left( \mu_i^{A}-\upsilon_i^A\right)\,, \qquad & i= {\nev/2}\,,\\
 \end{cases}
   \\ \nn
    \bar\phi^\cl_{i+\nev/2}\vert_{\Gamma_A} &= 
    \begin{cases}
 2 \pi R_c (M_{\nev/2-1})_{ij} \upsilon^A_j \,, \qquad & i, j=1, \dots, {\nev/2} -1\,,\\
 \sqrt{{\nev\over 2}} \chi_+ + {2 \pi \, R_c\over \sqrt{\nev}}\, \left(  I\cdot \mu^A + I\cdot \upsilon^A\right)\,, \qquad & i= {\nev/2}\,,\\
 %+ {2 \pi \, R_c\over \sqrt{\nev}}\,  \sum_{i=1}^{\nev/2-1}  \left( \mu_i^{A}+\upsilon_i^A\right)\,, \qquad & i= {\nev}\,.\\
 \end{cases}
\end{align}
\endgroup
\end{subequations}
where
$
I \cdot x : = \sum_{i=1}^{\nev/2-1} x_i \,
$
is the scalar product of $x\in  \Z^{\nev/2 -1}$ and $I$. 
Thus, the $\nev$--th field can be used to reconstruct the partition function over the whole manifold, while the $\nev/2$--th contributes to the partition function over the cylinder $A=A_1\cup A_2$. 
Hence, including the contributions from the fluctuating fields, our final expression is 
\begin{equation}
\label{tr-rho-A-partial-transpose-A2-torus-2}
\tr\big(\rho_{A}^{T_{A_2}}\big)^{n_e} = \sqrt{\nev} \, \frac{(Z_{A_1}Z_{A_2})^{n_e-2}Z_{A_1\cup A_2}Z_{B}^{n_e-1}}{Z_{A\cup B}^{n_e-1}}\, \WE(n_e)\,,
\end{equation}
where 
\begin{equation}
\label{general-def-we-torus}
\WE(\nev)=\sum_{\substack{\mu \in\Z^{\nev-2}\, \\  \upsilon \in\Z^{\nev-2}\,\\ \gamma \in \Z}}e^{-\sum_{k=1}^{\nev-1}S[\barphi^\cl_i]}\,,
\qquad \mu: = (\mu^1, \mu^A)\,, \quad \upsilon: = (\upsilon^1, \upsilon^A)\,, 
\end{equation}
and  the classical fields satisfy the boundary conditions \eqref{B-torus-new-notation-v3}.
Notice that  in \eqref{general-def-we-torus} the classical field $\barphi^\cl_{\nev/2}$ only sees two torus cuts,  along $\G_1$ and $\G_2$. 
As for the spherical geometry, the factor $\sqrt{\nev}$ appears in \eqref{tr-rho-A-partial-transpose-A2-torus-2} due to the different compactification radius for the $\nev$--th field. 
The logarithmic negativity \eqref{LNreplica} is then formally given by
\be
\label{curlye-t}
\E= -\log({Z_{A_1}Z_{A_2}\over Z_{A_1\cup A_2}}) + \log\WE(1)\,. 
\ee

The contribution from the fluctuating fields is straightforward to compute using the results reported in Appendix \ref{app:det-torus}%
\footnote{Again, the functional determinants in the above expressions are regularised  and calculated by means of zeta-function techniques.}
and  amounts to 
\be\label{det-torus-exp}
-\log(Z_{A_1} Z_{A_2}\over Z_{A_1\cup A_2})\, &=& {1\over 2} \log\left({\det \Delta_{A_1}\det \Delta_{A_2}\over \det\Delta_{A}}\right),
\nn\\
&=&{1\over 2} \log\left( {2 u_1 u_2\over u_{12}}|\tau|\right) + \log\left({|\eta (2 u_1 \tau ) \eta(2 u_2 \tau)|\over |\eta(2 u_{12} \tau)|}\right),
\ee
where we used the aspect ratios 
\be\label{aspect-ratios}
u_1={\ell_1\over L_1} \,, \qquad u_2={\ell_2\over L_1}\,, \qquad u_{12}=u_1+u_2\,,
\ee
for the submanifolds $A_1$ and $A_2$, respectively. 
Notice that 
$u_1 +u_2\neq 1$,
since there is still the contribution from the $B$ sector to the total lenght. 
This is even more pronounced in the winding function $\WE$. 
Following Appendix \ref{app:winding-torus}, the expressions for $\WE$ reads
\be
\label{we-torus-v2-2}
\WE(\nev)\nn\\\nn
&& \hspace{-35pt}=\sum_{\Z^{2\nev-3}} \exp\left\{- {g (2 \pi R_c)^2 \over \vert\tau\vert}\left[ 
\mu^T \, \mathrm T \, \mu +\upsilon^T \, \mathrm T \, \upsilon
+\left({1-\frac 2n}\right){1-u_2\over u_1 (1-u_{12})}\left(\gamma -I\cdot \mu^1\right)^2 \right.\right.
%+\left({1-\frac 2n}\right)\left({1\over u_1}+{1\over u_B}\right)\left(\gamma -I\cdot \mu^1\right)^2 \right.\right.
\\ \nn
&&%\hspace{-33pt}+\left.\left. \frac 1n\left(\frac{1}{u_B}+\frac{1}{u_1+u_2}\right)\left(\gamma -I \cdot \upsilon^1\right)^2 
\hspace{-33pt}+\left.\left.\frac{1}{n\, u_{12}(1-u_{12})}\left(\gamma -I \cdot \upsilon^1\right)^2 
%-{4\over n } \left(\gamma-I\cdot \mu^1 \right) \left( \left({1\over u_1}+{1\over u_B}\right)I\cdot \mu^1 - {1\over u_1}I\cdot \mu^A \right)\, 
-{4\over n\, u_1 } \left(\gamma-I\cdot \mu^1 \right) \left( {1-u_2\over u_1 (1-u_{12})}I\cdot \mu^1 - I\cdot \mu^A \right)\, 
%-{4\over n u_B} \gamma \left( I\cdot \mu^1\right)
\right]
\right\}, \\
\ee
where $\mathrm T$ is defined in \eqref{t-odd}. 
Collecting all the modes in a $(2\nev-3)$-vector, $\Omega = (\mu,\upsilon, \gamma)$,  we can rewrite the above expression \eqref{we-torus-v2-2} as follows 
\be
\label{we-torus-v4}
\WE(\nev) &=& \sum_{\Omega \in \Z^{2\nev-3}} \exp\left(- {g (2 \pi R_c)^2 \over \vert\tau\vert} \Omega^T \, \mathcal{T} \, \Omega\right),
\ee
where $\mathcal T$ is reported in Appendix \ref{app:winding-torus}, equation \eqref{huge-T}.  
The matrix $\mathcal T$ is symmetric and positive definite, hence the sum \eqref{we-torus-v4} is convergent. 
Unfortunately, we were not able to find an analytic continuation of the winding sector  \eqref{we-torus-v4}, which means that we are not able to compute $\log\WE(1)$ in \eqref{curlye-t}.

%%%%%%%%%%%%%%%%%%%%%%%%%%%%%%

%%%%%%%%%%%%%%%%%%%%%%%%%%%%%%%%%%%%%%%%%%%%%%%%%%%%%%%%%%%%
%%%%%%%%%%%%%%%%%%%%%%%%%%%%%%%%%%%%%%%%%%%%%%%%%%%%%%%%%%%%
\section{Odd entropy}
\label{sec:odd-EE}

The odd entropy $S_o$ introduced in~\cite{Tamaoka:2018ned} for a mixed state described by a density matrix $\rho_A$ is defined as 
\be\label{def-oee}
S_o^{(\nod)}(\rho_A)= {1\over 1- \nod} \left(\tr\big(\rho_A^{T_2}\big)^{\nod}-1\right) \,,\qquad  S_o(\rho_A)= \lim_{\nod \to 1 } S_o^{(\nod)}(\rho_A)\,,
\ee
where $\nod$ is an odd positive integer, and where, as before, we denote the union of $A_1$ and $A_2$ by $A$ and indicate the partial transposition over $A_2$ by $^{T_2}$. 
For pure states, the odd entropy reduces to the entanglement entropy \cite{Tamaoka:2018ned}, as per \eqref{nodd-pure}.

In this section we compute  $S_o$  by means of the replica approach. 
We stress that, as for the logarithmic negativity, our calculation only gives us the universal terms. 
The computations are similar to those illustrated before, and mainly differ from them in the boundary conditions at the entangling cut between $A_1$ and $A_2$, as we explain below. 
First, we consider a spherical manifold in Section \ref{sec:OEE-sphere}, and then we analyse the case of a toroidal manifold, see Section \ref{sec:OEE-torus}. 
On both geometries, we find the following formal expression for the odd entropy
\be
\label{oee-general}
S_o &=& -\log{ \left({Z_{A_1} Z_{A_2} Z_{B}\over Z_{A\cup B}}\right)}-\half -W'_{OE}(1) \,, 
\ee
where $W_{OE}$ is the contribution from the corresponding winding sector. 

%%%%%%%%%%%%%%%%%%%%%%%%%%%%%%%%%%
%%%%%%%%%%%%%%%%%%%%%%%%%%%%%%%%%%
\subsection{Spherical geometry}
\label{sec:OEE-sphere}

Let us consider a spherical geometry. In the next Section we discuss a geometrical configuration where the submanifolds $A_1$ and $A_2$ are disjoint, see Fig.~\ref{fig:realizations-of-disjoint-geometry}, and in Section \ref{sec:OEE-sphere-adjacent} we treat the case of adjacent submanifolds as in Fig.~\ref{fig:realizations-of-adjacent-geometry}. 
%
%

%%%%%%%%%%%%%%%%%%%%%%%%%%%%%%%%%%
\subsubsection{Disjoint submanifolds}
\label{sec:OEE-sphere-disjoint}

In the case of disjoints submanifolds and for an odd number of replicas, it is not difficult to realise that the gluing conditions force {\it all} the fields to agree at {\it both} entangling cuts. The situation is similar to the one depicted in Fig.~\ref{fig:boundary-conditions-of-disjoint-geometry}.
Using the same notation as in Fig.~\ref{fig:boundary-conditions-of-disjoint-geometry}, we now have the following boundary conditions 
\be
\Gamma_1:  &&\qquad \phi_i^{A_1}\vert_{\G_1}= \phi_j^{B}\vert_{\G_1} =\chi_1\,, \qquad i, j=1,\dots, \nod\,, \\ \nn
\Gamma_2:  &&\qquad \phi_i^{A_2}\vert_{\G_2}= \phi_j^{B}\vert_{\G_2} =\chi_2\,, \qquad i, j=1, \dots, \nod\,\,. 
\ee
Separating the classical contributions from the fluctuating fields, 
and using the global shift symmetry to eliminate the frequency modes from the entangling cut $\G_2$, 
we can write
\be
\Gamma_1:  &&\qquad \phi^{\rm cl}_i\vert_{\G_1}= \chi_1+2\pi R_c \hp \omega_i\,,   \qquad i=1,\dots, \nod -1\,, \nn\\
&&\qquad \phi^{\rm cl}_{\nod}\vert_{\G_1}= \chi_1\,, \\ \nn
\Gamma_2:  &&\qquad \phi^{\rm cl}_i\vert_{\G_2}= \chi_2\,, \hspace{2.5cm} i=1,\dots, \nod\,.
\ee
At this point we can perform the usual rotation $U_{\nod}$ and obtain 
\be\nn
\Gamma_1:  &&\qquad \tilde\phi^{\rm cl}_i\vert_{\G_1}= 2\pi R_c (M_{\nod-1})_{ij}\hp \omega_j,   \qquad i, j=1,\dots, \nod -1\,, \nn\\
&&\qquad \tilde\phi^{\rm cl}_{\nod}\vert_{\G_1}=\sqrt{\nod}\hp \chi_1 +{2\pi R_c\over \sqrt{\nod}} \sum_{i=1}^{\nod-1}\omega_i\,,  \\ 
\Gamma_2:  &&\qquad \tilde\phi^{\rm cl}_i\vert_{\G_2}= 0\,, \qquad\quad i=1,\dots, \nod-1\,, \nn\\
&&\qquad \tilde\phi^{\rm cl}_{\nod}\vert_{\G_2}=\sqrt{\nod}\,\chi_2 \,.\nn
\ee
The cassical field $\tilde\phi^{\rm cl}_{\nod}$ together with the fluctuating field $\varphi_{\nod}$ can be used to reconstruct the whole partition function over the sphere. 
Notice that we do not need to perform any further rotation. 
The $\nod-1$ classical fields $\tilde \phi_i^{\rm cl}$ contribute to the winding sector $W_{OE}(\nod)$ as before. 
The fluctuating fields obey Dirichlet conditions at the entangling cuts and give rise to the corresponding partition functions over the submanifolds. 
Hence, we can write
\be
\tr\big(\rho_A^{T_2}\big)^{\nod} = \sqrt{\nod} \left({Z_{A_1} Z_{A_2} Z_{B}\over Z_{A\cup B}}\right)^{\nod-1}W_{OE}(\nod)\,.
\ee
In particular, the expression for the winding sector $W_{OE}(\nod)$ is nothing but the function $W(\nod)$ given by \eqref{w-annulus}, where  $c$ now reads 
%where the constant $c$ in $W(\nod)$  \eqref{w-annulus} is
$$c={8 \pi^2 g R_c^2\over \log{(\eta_2/\eta_1)}}\,. $$
As in Section \ref{sec:sphere-logneg}, $\eta$ is the radial coordinate defined as $\eta=\tan{\theta/2}$, and 
the radii $\eta_{1(2)}$ correspond to the polar angles $\theta_{1(2)}$ where we place the cuts $\G_{1(2)}$. 
The odd entropy \eqref{def-oee} is then given by
\be\label{oee-sphere-initial}
S_o &=& -\log{ \left({Z_{A_1} Z_{A_2} Z_{B}\over Z_{A\cup B}}\right)}-\half -W'(1) \nn
\\
&=& \half \log{{\det \Delta_{A_1} \det \Delta_{A_2} \det \Delta_{B}\over \det \Delta_{\rm sphere}}}+\log{\sqrt{4\pi g \mathcal A} R_c}-\half -W'(1)\,,
\ee
which is nothing but the von Neumann entropy for two spherical caps $A_1, A_2$ computed in~\cite{Zhou:2016ykv}. 
The term $\log{\sqrt{4\pi g \mathcal A} R_c}$ is the contribution from the zero mode in the partition function on the sphere with area $\mathcal A$\cite{Zhou:2016ykv, Angel-Ramelli:2019nji}. 
The contribution from the regularised functional determinants (cf. Appendix \ref{app:det-sphere}) is
\be
&&\hspace{-5pt}\half \log{{\det \Delta_{A_1} \det \Delta_{A_2} \det \Delta_{B}\over \det \Delta_{\rm sphere}}}+\log{\sqrt{4\pi g \mathcal A} R_c}\nn
\\ \nn
&&\quad=\half \log{(\det\Delta_{\rm hemisphere})^2\over\det\Delta_{\rm sphere}}+\half \log \left({1\over \pi} \log{\eta_2\over \eta_1} \prod_{m>0}\left(1-\left({\eta_1\over \eta_2}\right)^{2m}\right)^2 \right)+\log{\sqrt{4\pi g \mathcal A} R_c} 
\\
&&\quad=\log{\sqrt{8\pi g} R_c}  +\half \log \left({1\over \pi} \log{\eta_2\over \eta_1} \prod_{m>0}\left(1-\left({\eta_1\over \eta_2}\right)^{2m}\right)^2 \right),
\ee
where we used that $\mathcal A=4\pi$ for the unit sphere and the result \eqref{det-sphere}. 
The contribution from the winding sector is~\cite{Zhou:2016ykv} 
\be
-W'(1)=-\half+\half\log{c}-\int_{-\infty}^{\infty} {dk\over \sqrt\pi}\,e^{-k^2}  \log \left( \sum_{\omega\in\Z} \exp\left(-{\pi\over c} \omega^2 -2 i \sqrt{{\pi\over c}}k \omega\right)\right). 
\ee
%where $c={8 \pi^2 R_c^2 g\over  \log{\eta_2\over \eta_A}}$. 
Finally, combining the two contributions, the odd entropy is given by
\be\label{oee-sphere-disjoint-final}
S_o &=& 2 \left(\log{\sqrt{8\pi g} R_c} -\half\right)+\log \prod_{m>0}\left(1-\left({\eta_1\over \eta_2}\right)^{2m}\right)^2 \nn \\
&&\qquad-\int_{-\infty}^{\infty} {dk\over \sqrt\pi}\,e^{-k^2}  \log \left( \sum_{\omega\in\Z} \exp\left(-{\pi\over c} \omega^2 -2 i \sqrt{{\pi\over c}}k \omega\right)\right). 
\ee
Summarising, for disjoint submanifolds, here represented by two spherical caps $A_1, A_2$, the odd entropy is identical to the entanglement entropy of $A_1\cup A_2$. 
For mixed states, one of the features of odd entropy is that it reduces to the corresponding von Neumann entropy if $\rho_A$ is a product state~\cite{Tamaoka:2018ned}. 
We remind the reader that for disjoint submanifolds the logarithmic negativity vanishes, see Sections \ref{sec:disjointLN} and \ref{sec:two-disjoint-intervals}. However, this is not a sufficient condition for a system to be unentangled. Hence, the results for the odd entropy and the logarithmic negativity are consistent.

%%%%%%%%%%%%%%%%%%%%%%%%%%%%%%%%%%
\subsubsection{Adjacent submanifolds}
\label{sec:OEE-sphere-adjacent}

Let us now consider the case in which $A_1$ and $A_2$ are adjacent as in Fig.~\ref{fig:realizations-of-adjacent-geometry}. 
Using Fig.~\ref{fig:gluing-adjacent} as a guiding example, but taking  an odd number of replicas, we can easily convince ourselves that {\it all} the fields have to agree at the entangling cuts, indicated here as $\G_A, \G_2$. 
We can then repeat the discussion of the previous section almost identically, with the replacement $\G_1\to \G_A$. 
It is thus clear that the odd entropy is given by \eqref{oee-sphere-initial}, where now the parameter $c$ in the function $W$ is given by 
$$
c={8 \pi^2 R_c^2 g\over  \log{(\eta_2/ \eta_A)}}\,,
$$
where $\eta_A= \tan{\theta_A/2}$ is the radial coordinate corresponding to the cut $\G_A$.%
\footnote{
Notice that now the odd entropy is formally identical to the von Neumann entropy for the two spherical caps $A_1$ and  $B$. 
}
Hence, the explicit expression of the odd entropy reads
\be\label{oee-sphere-adjacent-final}
S_o &=& 2 \left(\log{\sqrt{8\pi g} R_c} -\half\right)+\log \prod_{m>0}\left(1-\left({\eta_A\over \eta_2}\right)^{2m}\right)^2 \nn\\
&&\qquad-\int_{-\infty}^{\infty} {dk\over \sqrt\pi}\,e^{-k^2}  \log \left( \sum_{\omega\in\Z} \exp\left(-{\pi\over c} \omega^2 -2 i \sqrt{{\pi\over c}}k \omega\right)\right).
\ee

%%%%%%%%%%%%%%

In the case of adjacent submanifolds, we can examine the pure state limit, as we have done for the logarithmic negativity. 
There are, however, some caveats. 
Looking at the expression \eqref{oee-general}, the divergences which we can expect are given by the interfaces between $A_1$ and $A_2$, and between $A_2$ and $B$ \cite{Cardy:1988tk}, see also discussion in~\cite{Zhou:2016ykv}.
In the pure state limit, the odd entropy is expected to reduce to the von Neumann entropy of the corresponding pure state, which has a divergence controlled by the characteristic size of the entangling surface $\G_A$ between $A_1$ and $A_2$. 
Since we are only computing the universal terms, expression \eqref{oee-general} also develops a divergent contribution related to the entangling surface $\G_2$ between $A_2$ and $B$ in the pure state limit. 
This means that in our approach only a ``regulated'' odd entropy, given by the difference of the odd entropy and an entanglement measure carrying a divergence at the entangling surface $\G_2$, will correctly reduce to the entanglement entropy in the pure state limit. 
We choose the entanglement entropy $S_{EE}(\rho_A)$ of the region $A$ as regulator, since it carries the divergence that we want to subtract. 
In this case, $S_{EE}(\rho_A)$ is the entanglement entropy of a spherical cap given by the union of $A_1, A_2$, that is~\cite{Zhou:2016ykv, Angel-Ramelli:2019nji}
\be
S_{EE}(\rho_A)= \frac 12 \log \left( {\det \Delta_A \det \Delta_B \over \det_{A\cup B}}\right)+\log\left(\sqrt{4\pi g \mathcal A} R_c\right)-\frac 12= \log\left(\sqrt{8\pi g} R_c\right)-\frac 12\,. ~~~
\ee
Hence, in the pure state limit, when $\eta_2\gg \eta_A$, that is when $c\ll1$, we have 
\be\label{OE-pure-limit-reg}
\Delta S_o \equiv S_o -S_{EE} (\rho_A)  &\approx& \log{\sqrt{8\pi g} R_c} -\half-2 \left({\eta_A\over \eta_2}\right)^{\frac{1}{4 \pi  g R_c^2}}- \left({\eta_A\over \eta_2}\right)^2+\dots\,,
\ee
which correctly reproduces the pure state entanglement entropy result~\cite{Zhou:2016ykv}.
We have checked that using $S_{EE}(\rho_A)$ as a regulator also provides a correct pure state limit in the case of cylindrical manifolds (in the non-compact case). 
There, the term removed by the entanglement entropy is actually divergent and $\Delta S_o$ correctly reduces to the finite universal term of the entanglement entropy for a pure state, cf. Appendix \ref{app:OE}.

%%%%%%%%%%%%%%%%%%%%%%%%%%%%%%%%%%%%%%%%%%%%%%%%%%%%%
%%%%%%%%%%%%%%%%%%%%%%%%%%%%%%%%%%%%%%%%%%%%%%%%%%%%%
\subsection{Toroidal geometry}
\label{sec:OEE-torus}

Next, we consider the toroidal geometry depicted in Fig.~\ref{fig:realizations-of-adjacent-geometry}. 
%We remind the reader that we first trace over $B$, and then we partially transpose over $A_2$. 
Let us proceed as in Section \ref{sec:torus-logneg} in order to understand the boundary conditions at each entangling cuts. 
When the replica index $n=\nod$ is odd, the boundary conditions at the cut $\G_A$ between $A_1$ and $A_2$ require {\it all} fields to be equal,%
\footnote{We remind the reader that for an even integer $\nev$ the boundary conditions at $\G_A$ split the fields into two independent sets, cf. Section \ref{sec:torus-logneg}.}
while the conditions at the cuts $\G_1$ and $\G_2$ remain unchanged. 
After separating each of the fields into a fluctuation $\varphi_i$ satisfying Dirichlet boundary conditions at the cuts and a classical part $\phi^\cl_i$, the boundary conditions for the classical fields at the entangling cuts read
\be
&&\phi^\cl_i\vert_{\G_a} =  \chi_a +2 \pi\, R_c \, \omega_i^{a}  +\phi_i^0-\phi_{\nod}^0\,, \qquad i=1, \dots, \nod-1\,,\\ \nn
&&\phi^\cl_{\nod}\vert_{\G_a} =  \chi_a \,, 
\ee
for  $a=1, 2, A$, where $\phi_i^0-\phi_n^0$ are the zero modes. Note that we used our freedom to redefine the arbitrary cut functions $\chi_a$ to get rid of the $n_o$--th winding numbers and zero mode at all cuts, and relabelled $\omega_i^a -\omega_{\nod}^a$ to $\omega_i^a$ without loss of generality. 
We can choose $\phi_i^0-\phi_{\nod}^0$ such that it removes the winding modes at the entangling cut $\G_2$, the choice of $\G_2$ here is completely equivalent to any other, since all the entangling cuts carry the same number of degrees of freedom. 
With this choice we obtain  
\be
&&\phi^\cl_i\vert_{\G_a} =  \chi_a +2 \pi\, R_c \, \left( \omega_i^{a} - \omega_i^{2}\right), \qquad a=1, A\,, \quad i=1, \dots, \nod-1\,,\nn\\
&&\phi^\cl_{\nod}\vert_{\G_a} =  \chi_a\,, \hspace{3.94cm} a=1, A\,, 
\\ \nn
&&\phi^\cl_{i}\vert_{\G_2} =  \chi_2\,,  \hspace{4.1cm}  i\hp=1, \dots, \nod\,\,. 
\ee
We can then perform a $U_{\nod}$ rotation, cf. \eqref{def-matrix-U}, which gives us
\be
\label{bc-woee-torus}
&&\tilde \phi^\cl_i\vert_{\G_a} = 2 \pi\, R_c \, \left(M_{\nod -1}\right)_{ij}\omega^{a}_i,  \hspace{1.61cm} a=1, A\,, \quad i=1, \dots, \nod-1\,,\nn\\
&&\tilde \phi^\cl_{\nod}\vert_{\G_a} = \sqrt{\nod}  \chi_a+{2\pi \, R_c\over \sqrt{\nod}} \sum_{k=1}^{\nod-1} \omega_i^{a}\,, \qquad a=1, A\,, 
\\ \nn
&&\tilde \phi^\cl_{i}\vert_{\G_2} =0\,, \hspace{4.6cm}  i\hp=1, \dots, \nod-1\,,\\
&&\tilde \phi^\cl_{\nod}\vert_{\G_2}=\sqrt{\nod}\,  \chi_2 \,, \nn
\ee
where we, again,  relabelled $\omega_i^{a} - \omega_i^{2}$ into $ \omega^{a}_i$ without loss of generality. 
Notice that we have in total $2\nod -2$ independent winding modes $\omega^{a}_i$, which we can collect in a $\Z^{2\nod -2}$ vector as $\omega: = (\omega^{1}, \omega^{A})$. 
As usual, the $\nod$--th field can be used to reconstruct the partition function on the entire torus $Z_{A\cup B}$, and we obtain
\be\label{expression-tr}
\tr\big(\rho_A^{T_2}\big)^{\nod} = \sqrt{\nod} \left({Z_{A_1} Z_{A_2} Z_{B}\over Z_{A\cup B}}\right)^{\nod-1}W_{OE}(\nod)\,,
\ee
where $W_{OE}(\nod)$ is again the contribution from the $2\nod-2$ classical modes, now satisfying the boundary conditions \eqref{bc-woee-torus}, that is 
\be\label{oee-winding-torus-gen}
W_{OE}(\nod) =\sum_{\phi^\cl_i} e^{-\sum_i S[\phi^\cl_i]}\,. 
\ee
Hence, the formal expression for the odd entropy \eqref{def-oee} which follows from \eqref{expression-tr} is
\be
\label{oee-torus-semifinal}
S_o &=& -\log{ \left({Z_{A_1} Z_{A_2} Z_{B}\over Z_{A\cup B}}\right)}-\half -W'_{OE}(1) \,. 
\ee

The contribution from the fluctuating fields can be read straightforwardly from the results reported in Appendix \ref{app:det-torus}, and we obtain 
\be
\nn
-\log{ \left({Z_{A_1} Z_{A_2} Z_{B}\over Z_{A\cup B}}\right)}&=& \half \log{{\det \Delta_{\rm cyl, A_1} \det \Delta_{\rm cyl, A_2} \det \Delta_{\rm cyl, B} \over \det \Delta_{\rm torus}}} +\half \log{4 \pi g R_c^2 \mathcal A} \qquad  \\\nn
 &=& \half \log \left({8 u_1 u_2 (1-u_{12}) \vert \tau\vert^2}\right)+ \log {{\eta(2 u_1\tau ) \eta(2 u_2\tau ) \eta(2 (1-u_{12})\tau )\over \eta^2(\tau )}}
 \\ 
 && +\half \log{4 \pi g R_c^2}\,,\label{oee-torus-det-explicit}
 \ee
 since $\mathcal A=L_1 L_2$, and where the Dedekind $\eta$-function is defined in \eqref{def-dedekind-eta}. The aspect ratios where defined in \eqref{aspect-ratios}.

%%%
The explicit expression for the winding sector is calculated in Appendix \ref{app-winding-sector-oe}, see equations \eqref{winding-torus-oee} and \eqref{eq:appendix-winding-OE-final}, and reads 
\be\label{sec-4-winding-torus-oee-final}
W_{OE}(\nod) = 
\nod \, {\left({ |\tau|\over 4 \pi g R_c^2} \right)^{(\nod-1)}} \left({ u_1 u_2 (1-u_{12})}\right)^{(\nod-1)/2} 
\Theta(\vec 0\,\vert \,\mathrm{U})\,,
\ee
where $\Theta(\vec{0}\,\vert\mathrm U)$ is a multi-dimensional theta function and $\mathrm U$ a positive definite matrix. Unfortunately, we have not found an analytic continuation to real $\nod$ for the winding sector \eqref{sec-4-winding-torus-oee-final}, hence we cannot compute the derivative of  $W_{OE}$ at $\nod=1$. 

%%%%%%%%%%%%%%%%%%%%%%%%%%%%%%%%%%%%%%%%%%%%%%%%%%%%%%%%%%%%%%%%%%%%%%%%%%%%%%%%%%%%%%%%%%%%%%%%%%%%%%%%%%%%%%%%%%%%%%%%%%%%%%%%%%%%%%%%%%%%%%%%%%%%%%%%%%%%%%%%%%%%%%%%%%%%%%%%%%%%%%%%%%%%%%%%%%%%%%%%%%%%%%%%%%%%%%%%%%%%%%%%%%%%%%%%%%%%%%%%%%%%%%%%%%%%%%%%%%%%%%%%%%%%%%%%%%%%%%%%%%%%%%%%%%%%%%%%%%%%%%%%%%%%%%%%%%%%%%%%%%%%%%%%%%%%%%%%%%%%%%%%%%%%%%%%%%%%%%%%%%%%%%%%%%%%%%%%%%%%%%%%%%%%

\section{Discussion}
\label{sec:discussion}

In this paper, we computed the logarithmic negativity for the quantum Lifshitz model -- a prototype of non-relativistic field theories described by a free massless compact scalar with Lifshitz exponent $z=2$ -- in one and two spatial dimensions. 
To this end, we employed two different techniques: the correlator method for the $(1+1)$-dimensional QLM, where we assumed the scalar  to be non-compact, and a replica method in $2+1$ dimensions.  
In both cases, we first examined the QLM in its ground state and confirmed that the logarithmic negativity of a pure state reduces to the R\'enyi entropy with index $1/2$, as generically expected in QFT~\cite{Calabrese2013}. 
We then investigated the QLM in a bipartite mixed state ($A_1, A_2$) obtained by tracing out the degrees of freedom of some subsystem (referred to as $B$ throughout the paper) of the tripartite ground state.
In particular, we studied the two cases where the subsystems $A_1$ and $A_2$ are either disjoint or adjacent. 
Both methods lead to the same general result: 
the logarithmic negativity vanishes for disjoint partitions regardless of the manifold, 
while it is given by a difference of free energies for adjacent subsystems.

A common feature we observed is the feasibility of analytic computations. 
As already mentioned, in one spatial dimension, the moments of the reduced density matrix and its partial transpose are obtained analytically, something that is not yet possible in $(1+1)$-dimensional CFT. 
Moreover, in the $(2+1)$-dimensional case, the computation of the logarithmic negativity simply reduces to a computation of partition functions for a free scalar relativistic theory, albeit on a non-trivial geometry. 
This is very different from what happens in CFTs, see e.g.~\cite{Calabrese:2012ew, Calabrese2013}, and the feasibility of calculations in the QLM, in comparison with the conformal paradigm, is somehow surprising. 

Another unexpected result is the fact that for disjoint submanifolds the QLM has a vanishing logarithmic negativity, similar to a topological theory, while for contiguous submanifolds the ``CFT character'' of the QLM dominates. 
We should stress here that the expression for the logarithmic negativity \eqref{LNgen} in the $(1+1)$-dimensional QLM is suggestive of a relativistic field theory but on an infinite system (see discussion in Section \ref{sec:hint-at-a-general-formula}) and with an effective central charge given by $c_{\rm eff}= z c_{\rm CFT}=2$, as proposed in~\cite{MohammadiMozaffar:2017chk}. 

To be more specific, the calculations in the real time formalism are carried out for a non-compact scalar on open and periodic chains.
In the Euclidean formalism, the scalar is periodically identified (with the exception of section \ref{sec:adjacent-no-winding} which we 
use as a ``warm-up exercise'') and we take the spatial manifold to be either a 2-sphere or a 2-torus. 
For adjacent submanifolds on the sphere, we are able to analytically continue the winding sector contribution $\WE$, but
regrettably,  we have not found a corresponding analytic continuation for $\WE$ on the torus. 
The crucial difference between the two geometries, rendering one case completely solvable and the other not, is that only 
one entangling cut is ``visible'' to the winding sector on the sphere, while there are two ``visible'' cuts on the torus, see Sections \ref{sec:sphere-logneg} and \ref{sec:torus-logneg}. 

We also computed the odd entropy for the QLM on a 2-sphere and a 2-torus by means of the replica method. 
For the spherical case, regardless of whether the submanifolds are adjacent or disjoint, we always find a non-trivial result.
Furthermore, the expressions for the odd entropy coincide with that of the entanglement entropy for two spherical caps, 
originally computed in~\cite{Zhou:2016ykv}. 
This confirms the expectation that the odd entropy encodes both classical and quantum fluctuations~\cite{Tamaoka:2018ned}.  
For a toroidal manifold, we were not able to analytically continue the expression for the winding sector contribution to the odd entropy. 

Interestingly, we notice that for non-compact fields the following relation holds:
\be
S_o(\rho_A) - S_{EE}(\rho_A) = \E(\rho_A)\,, \lb{relation}
\ee
where the odd entropy is given by keeping  only the first term in \eqref{oee-general}, the entanglement entropy is formally given by \eqref{fradk} with $A=A_1\cup A_2$, and the logarithmic negativity can be found in \eqref{curly-E-gen-replica-nowinding}.
For holographic CFTs, the quantity $S_o -S_{EE}$ is conjectured \cite{Tamaoka:2018ned} to be equal to the entanglement wedge cross-section (EWCS) in AdS spacetime, while it has been proposed%
\footnote{Another proposal for the holographic dual of entanglement negativity has been suggested in \cite{Chaturvedi:2016rcn,Jain:2017sct,Jain:2017xsu}, and relates the logarithmic negativity in holographic CFTs to a certain combination of bulk minimal surfaces which reduces to the holographic mutual information.}
 in \cite{Kudler-Flam:2018qjo,Kusuki:2019zsp} that the logarithmic negativity $\E$ should to be dual to a backreacted EWCS. In particular, for simple subsystem configurations in $2d$ holographic CFTs, the backreacted EWCS picks up a factor of $3/2$ compared to that computed in pure AdS$_3$. We thus observe through \eqref{relation} a clear difference between Lifshitz and conformal field theories. Let us also point out that \eqref{relation} breaks down for compact fields -- the relation being spoiled by the winding sector.

To our knowledge, our findings represent the first analytical results for the logarithmic negativity and the odd entropy in $2+1$ dimensions. Numerical studies for the logarithmic negativity in 3$d$ CFTs were conducted in~\cite{PhysRevB.93.115148,DeNobili:2016nmj}. 
This is one of the main remarkable properties of the QLM (and its higher-dimensional generalisations): it is a solvable theory for which one can perform controlled calculations in closed form, thus allowing us to extend analytical techniques beyond the 2$d$ conformal framework, and also providing a benchmark for numerical investigations. 

A substantial part of this work was focused on spherical and toroidal manifolds in $2+1$ dimensions. 
It would be interesting to consider other geometries, such as disks, and more general partitions. 
However, increasing the number of entangling cuts considerably increases the difficulty in performing analytic 
continuations. 
Indeed, the analytic continuation $\nev\to 1$ of the winding mode contributions remains an open problem (the same difficulties are present in 2$d$ CFTs, see~\cite{Calabrese:2012ew, Calabrese2013}). 
The calculation of other entanglement measures for mixed states of relativistic and non-relativistic systems remains a 
relevant and challenging open problem. %

%%%%%%%%%%%%%%%%%%%%%%%%%%%%%%%%%
%%%%%%%%%%%%%%%%%%%%%%%%%%%%%%%%%
\section*{Acknowledgments}

We thank B. Chen, N. Jokela, K. Tamaoka, and especially D. Seminara and E. Tonni for valuable discussions and comments. 
VGMP would like to thank the Nordic Institute for Theoretical Physics (NORDITA), the University of Edinburgh, and the Galileo Galilei Institute for Theoretical Physics in Florence for the hospitality during the completion of this work. 
This research was supported in part by the Icelandic Research Fund under contracts 163419-053 and 195970-051, and by grants from the University of Iceland Research Fund. C.B. was supported in part by the National Natural Science Foundation of China (NSFC, Nos. 11335012, 11325522, 11735001), and by a Boya Postdoctoral Fellowship at Peking University.

%%%%%%%%%%%%%%%%%%%%%%%%%%%%%%%%%
%%%%%%%%%%%%%%%%%%%%%%%%%%%%%%%%%
\appendix

\addtocontents{toc}{\protect\setcounter{tocdepth}{1}}

\section{Invariance of the reduced density matrix under partial transposition for two disjoint intervals}
\lb{Apdx}

In this appendix, we show that the reduced density matrix for two disjoint intervals in the $z=2$ open chain is invariant under partial transposition. 
In the position representation, the ground state wave function reads
\be
\Psi_0(\phi) \propto \exp\Big(\hspace{-3pt}-\frac{1}2 \phi^T W \phi\Big)\,,
\ee
and the associated density matrix is
\be
\rho(\phi,\phi') \propto \exp\Big(\hspace{-3pt}-\frac{1}2(\phi^T W \phi+ \phi'^T W \phi')\Big)\,,
\ee
where $\phi^T=(\phi_1 ,\phi_2,\cdots)$, and $W\equiv K^{1/2}$ with the matrix $K$ being a discrete version of the spatial biharmonic operator $\triangle^2\equiv\partial^4_x$, see Section \ref{sec:QLM}.

For simplicity, we consider $A=A_1\cup A_2$ to be a subsystem of the $z = 2$ open chain with $A_1$ and $A_2$ disjoint (of lengths $\ell_1$ and $\ell_2$, respectively) and each adjacent to one of the boundaries of the total system, although our reasoning carries through to general disjoint configurations. 
The reduced density matrix corresponding to $A$ is obtained by integrating over the degrees of freedom in $B$. To perform the integration explicitly, we express $\phi = (\phi_{A_1}^T,\phi_B^T,\phi_{A_2}^T)^T$ in terms of oscillators belonging to each subsystems, and break $W$ down into submatrices
\be
W=
\left(\begin{array}{ccc}
W_{A_1} & V_1^T & V_A \\
V_1 & W_B & V_2 \\
V_A^T & V_2^T & W_{A_2}
\end{array}\right).
\ee
It is then straightforward to obtain the reduced density matrix for the subsystem $A$ as
\be
\rho_A(\phi_A,\phi_A') \propto  \exp\Big(\hspace{-3pt}-\frac{1}2 \overline{\phi}^T_A\,\overline{W}\,\overline{\phi}_A \Big)\,,
\ee
where $\overline{\phi}_A = (\phi_{A_1}^T,\phi_{A_2}^T,\phi_{A_1}^{'T},\phi_{A_2}^{'T})^T$ and
\be
\overline{W}=
\left(\begin{array}{cccc}
\overline{W}_{\hspace{-3pt}A_1} & \overline{V}_{\hspace{-3pt}A} & \overline{V}_{\hspace{-2pt}11} & \overline{V}_{\hspace{-2pt}12}\\
\overline{V}_{\hspace{-3pt}A}^T & \overline{W}_{\hspace{-3pt}A_2} &  \overline{V}_{\hspace{-2pt}12}^T & \overline{V}_{\hspace{-3pt}22} \\
 \overline{V}_{\hspace{-2pt}11} &  \overline{V}_{\hspace{-2pt}12} & \overline{W}_{\hspace{-3pt}A_1} & \overline{V}_{\hspace{-3pt}A}  \\
\overline{V}_{\hspace{-2pt}12}^T &  \overline{V}_{\hspace{-2pt}22} & \overline{V}_{\hspace{-3pt}A}^T & \overline{W}_{\hspace{-3pt}A_2}
\end{array}\right), \qquad\qquad
\begin{aligned}
&\overline{W}_{\hspace{-3pt}A_a} = W_{A_a}+ \overline{V}_{\hspace{-2pt}aa}\,,\\
&\overline{V}_{\hspace{-2pt}ab} = -\frac{1}{4} V_a^TW_B^{-1}V_b\,, \quad a,b=\{1,2\}\,,\\
&\overline{V}_{\hspace{-3pt}A} =  V_A +  \overline{V}_{\hspace{-2pt}12}\,.
\end{aligned}
\ee
Invariance under partial transposition of the reduced density matrix, that is $\rho_A^{T_a}=\rho_A$, is equivalent to the condition $V_A=\mathbf{0}_{\ell_1\times\ell_2}$. We have checked%
\footnote{The matrix $W$ is nothing else than (twice) the two-point function given explicitly in \eqref{Pcorr}, where one can readily check that $V_A$ is a zero matrix.} 
 that this condition always holds for two disjoint intervals in the critical $z=2$ open chain, thus implying the invariance of $\rho_A$ under partial transposition.

\section{Spectrum of $C_A^{T_2}$ for two adjacent intervals} 
\lb{Apdx1}

\paragraph{Open system.}
For two adjacent intervals of arbitrary lengths and relative position in the total system (say, $A_1$ is at a distance $d_l$ from the left boundary), there are at most four eigenvalues of $C_{A}^{T_2}$ distinct from $1/2$. They are the (positive) roots of the quartic equation $a \lambda^8 + b\lambda^6 + c\lambda^4 + d\lambda^2 +e=0$ where
\be
a&=& 256(L+1)\,,\\
b&=&-64\big((L-2d_l)(5\ell_1+\ell_2+6d_l+4)-(\ell_1+\ell_2)^2-4(\ell_1^2-1) +2d_l(3d_l+4) \big)\,,\\
c&=& 16\big(6 - 7 \ell_1^2 - 2 \ell_1 (1 + 2 \ell_1) \ell_2 - (3 + 4 \ell_1) \ell_2^2 - 
 5 d_l^2 (1 + \ell_1 + \ell_2)  \,,\nonumber\\
 &&+ L (6 + 7 \ell_1 + 3 \ell_2 + 4 \ell_1 \ell_2)- d_l (9 \ell_1 + \ell_2 + 5 (\ell_1 + \ell_2)^2 - 5 L (1 + \ell_1 + \ell_2)) \big)\,,\\
d&=&4 \big( 3 \ell_1^2 -4+ 2 \ell_1 (1 + 2 \ell_1) \ell_2 + (3 + 4 \ell_1) \ell_2^2 + 
   2 d_l^2 (2 + \ell_1 + \ell_2 + 2 \ell_1 \ell_2) \,,\nonumber\\
   &&- 2 d_l (L - \ell_1 - \ell_2) (2 + \ell_1 + \ell_2 + 2 \ell_1 \ell_2) - L (4 + 3 \ell_1 + 3 \ell_2 + 4 \ell_1 \ell_2)\big) \,,\\
e&=& (d_l+1) (\ell_1 + \ell_2+1) (L - \ell_1 - \ell_2-d_l+1)\,.
\ee
These roots are real and positive. Among them, only one is smaller than $1/2$. It is given by
\be
\lambda^2 = -\frac{b}{4a}- S-\frac{1}{2}\sqrt{\frac{p}{S}-2q-4S^2}\,,
\ee
where
\be
p&=& \frac{b^3-4abc+8a^2d}{8a^3}\,, \quad q=\frac{8ac-3b^2}{8a^2}\,, \quad S=\sqrt{(\Delta_0\cos(\phi/3)-a p)/(6a)}\,,\\
\Delta_0&=&\sqrt{c^2-3bd+12ae}\,, \quad \phi=\arccos\Big(\frac{\Delta_1}{2\Delta_0^3}\Big)\,,\\
\Delta_1&=&2c^3-9bcd+27b^2e+27ad^2-72ace\,.
\ee
This eigenvalue $\lambda$ may also be expressed with radicals, but its form is far too cumbersome to be displayed here.
Now, considering the continuum regime, we find that $\lambda$ does not actually depend on $L$ nor $d_l$, i.e.
\be
\lambda\, \stackrel{\rm cont.}{\longrightarrow}\, \sqrt{\frac{\ell_1+\ell_2}{16\ell_1\ell_2}}\,.
\ee
This may indeed be checked numerically.

\paragraph{Periodic system.}

For the general case, with arbitrary $\ell_1+\ell_2<L$, the eigenvalues of $C_{A}^{T_2}$ in the critical limit ($m\rightarrow 0$, $m L\ll1$) are
\be
\spec(C_A^{T_2}) = \left\{\lambda_1,\lambda_2,\lambda_3,\sqrt{\frac{3}{2m^2L}},\frac{1}{2},\,\cdots,\frac{1}{2}  \right\}\,,\lb{eigPM}
\ee
where $\lambda_{1,2,3}$ are the roots of the cubic equation $a \lambda^6- b \lambda^4+c\lambda^2 -d=0$ with
\be
a&=& 384L\,,\\
b&=& 16\big(5(L-\ell_1-\ell_2+1)(\ell_1+\ell_2+1)+8\ell_1\ell_2+4\big)\,,\\
c&=& 8\big((L-\ell_1-\ell_2+1)(\ell_1+\ell_2+2\ell_1\ell_2+2)+\ell_1+\ell_2+1\big) \,,\\
d&=& (L-\ell_1-\ell_2+1)(\ell_1+\ell_2+1) \,.
\ee
Only one eigenvalue in the spectrum of $C_A^{T_2}$, among $\lambda_{1,2,3}$, is smaller than $1/2$ and  in the continuum limit it reads \mbox{$\lambda^2=(\ell_1+\ell_2)/(16\ell_1\ell_2)$}. Thus, similarly to the open chain, $\lambda$ does not depend on the size of the total system. 

%%%%%%%%%%%%%%%%%%%%%%%%%%%%%%%%%%%%%%%
%%%%%%%%%%%%%%%%%%%%%%%%%%%%%%%%%%%%%%%
\section{Functional determinants and reciprocal formulae}
\label{app:det}

%%%%%%%%%%%%%%%%%%%%%%%%%%%%%%%%%%%%%%%
\subsection{Spherical manifolds}
\label{app:det-sphere}

Here we list the results for the various  regularised functional determinants computed by means of zeta-function regularisation techniques in~\cite{Weisberger:1987kh, Dowker:1993hq, Dowker:1995sp}, and relevant for the spherical case discussed in Section \ref{sec:adjacent-manifolds-winding}. 
In the following, whenever the manifold has a boundary, Dirichlet boundary conditions are assumed. 
\be
&&\half \log\det \Delta_{\rm spherical ~cap}=\half \log \det \Delta_{\rm hemisphere} -{1\over 3} \cos\theta -{1\over 6} \log\tan{\theta\over 2}\,, \nn\\
&&\half \log\det \Delta_{\rm between ~ spherical ~caps}=\half \log \det \Delta_{\rm annulus} +{1\over 3} \left( \cos\theta_{\rm in} -\cos\theta_{\rm fin}\right)\,, \\\nn
&&\det \Delta_{\rm annulus} = {1\over \pi} \, \mu^{1/3}\, \log {1\over \mu}\,  \prod_{m>0}(1-\mu^{2m})^2 \,,  \qquad \mu={\eta_{\rm int}\over \eta_{\rm out}}\,. 
\ee
The angles $\theta_{\rm in}, \theta_{\rm fin}$ are the smaller and larger angles respectively, starting from the North pole, which delimit the spherical surface between two spherical caps. 
The radii $\eta_{\rm int}, \eta_{\rm out}$ are the internal and external radii respectively of the annulus. 
The radial coordinate $\eta$ and the polar angle $\theta$ are related by a stereographic projection, that is 
\be
\nn
\eta= \tan{{\theta\over 2}}\,. 
\ee
Finally, the functional determinants for the Laplacian on the sphere and hemisphere are
\be
\label{det-sphere}
&&\det \Delta_{\rm hemisphere}=\frac{e^{\frac{1}{4}-2 \left(\frac{1}{12}-\log (A)\right)}}{\sqrt{2 \pi }}\,,
\nn\\ 
&&\det \Delta_{\rm sphere}=e^{\frac{1}{2}-4 \left(\frac{1}{12}-\log (A)\right)}\,,
\\ \nn
&&\half \log{(\det\Delta_{\rm hemisphere})^2\over\det\Delta_{\rm sphere}}={1\over 2}\log{1\over 2\pi}\,,
\ee
where $A$ is the Glaisher constant.

%%%%%%%%%%%%%%%%%%%%%%%%%%%%%%%%%%%%%%%
\subsection{Toroidal manifolds}
\label{app:det-torus}

Here we report the results for the regularised functional determinants of the Laplacian operator on a cylinder and a torus~\cite{DiFrancesco1997, Ginsparg:1988ui}, see also \cite{Angel-Ramelli:2019nji} and references therein for higher-dimensional generalisations.
Consider a two-dimensional torus with area $L_1\times L_2$. The functional determinant of the Laplacian on the torus is
\be
\label{logdet-torus}
\log\det \Delta_{\rm torus}=\log(L_1^2 \, \eta^4(\tau))\,, \qquad \tau=i {L_1 \over L_2}\,,
\ee
where the Dedekind $\eta$-function is defined as follows
\be\label{def-dedekind-eta}
\eta(\tau ):= q^{1/24} \prod_{n=1}^\infty (1-q^n)\,, \qquad q:=e^{2 i\pi \tau}\,. 
\ee
Notice that the partition function on the torus is given by
\be
Z_{\rm torus}= 2 \pi R_c {\sqrt{{g  \mathcal A\over \pi}}} {\det}^{-\half} \Delta_{\rm torus}\,,
\ee
due to the presence of a zero mode, with $\mathcal A=L_1L_2$ the area of the torus. 

The functional determinant of the Laplacian operator on a cylinder of length $L$, that is $[0,L]\times S^1_{L_2}$, with Dirichlet boundary conditions, is
\be\label{det-cyl}
\log \det \Delta_{\rm cylinder}= \log( 2 \alpha |\tau| \eta^2(2 \alpha \tau))\,, \qquad  \alpha= {L\over L_1} \,, ~~\tau=i {L_1 \over L_2}\,. 
\ee

%%%%%%%%%%%%%%%%%%%%%%%%%%%%%
\subsection{Reciprocal formulae}
\label{app:formulae}

The reciprocal formula for the theta function is 
\be
\label{reciprocal-formula}
\sum_{\omega\in\Z} \exp\hspace{-1pt}\left(-{\pi\over c} \omega^2 -2 i \sqrt{{\pi\over c}}k \omega\right)=\sqrt{c}\,e^{-k^2}\sum_{\omega\in\Z} \exp\hspace{-1pt}\left(-{\pi c\hp \omega^2} + 2\sqrt{{\pi c}}\,k \omega\right). 
\ee

\noindent
The reciprocal formula for the multi-dimensional theta function~\cite{MR136775} is
\be
\label{reciprocal-formula-multi}
\sum_{\vec m \in \Z^{n}} \exp\left( - \pi\hp \vec m^T \hspace{-1.5pt}A\hp \vec m+2i \pi \hp\vec m\cdot\vec z\right)= {\det A}^{-\half} \sum_{\vec p\in \Z^{n}}\exp\left( - \pi(\vec p+\vec z)^T\hspace{-1.5pt} A^{-1}(\vec p+\vec z)\right). ~~~
\ee

%%%%%%%%%%%%%%%%%%%%%%%%%%%%%
%%%%%%%%%%%%%%%%%%%%%%%%%%%%%
\section{Winding sector for the 2-torus}
\label{app:winding-torus}

In this section we illustrate in some detail the computation of the winding sectors $\WE$ and $W_{OE}$ for the torus, see equations \eqref{general-def-we-torus} and \eqref{oee-winding-torus-gen}. 
In order to construct the factor $\sum_{k=1}^{\nev-1}S[\barphi^\cl_i]$ appearing in $\WE$ and $W_{OE}$, we need to find the classical solutions $\barphi^\cl_i$ which satisfy the corresponding boundary conditions, that is  \eqref{B-torus-new-notation-v3} and \eqref{bc-woee-torus}, respectively. Due to our surgery procedure, the classical fields do not depend on the direction along the cuts~\cite{Zhou:2016ykv, Angel-Ramelli:2019nji}, and we only have to solve the equations of motion on an interval.
We remind the reader that  a generic solution to the Laplace boundary value problem on the interval $[a, b]$, that is 
$$
\p_x \p^x f(x)=0\,, \qquad f(a)=f_a\,, \qquad f(b)=f_b\,,\qquad x\in [a,b]\,, 
$$
is given by
$$
f(x)= {f_b-f_a\over b-a} x+ {f_a b-f_b a\over b-a}\,.
$$
In the winding sectors $\WE$ and $W_{OE}$, the only non-trivial term entering is $\p_x f \p_x f$, and it brings down the factor $\big(\frac{f_b-f_a}{b-a}\big)^2$.

%%%%%%%%%%%%%%%%%%%%%%%%%%%%%
\subsection{Winding sector $\WE$}

In this section, we illustrate the computation of the winding sector $\WE$ which appears in the logarithmic negativity for a toroidal manifold \eqref{general-def-we-torus}. 
The classical fields $\barphi^\cl_i$ for $i=1, \dots, \nev-1$ and $i\neq\nev/2$ have support on the torus  of area $L_1\times L_2$ divided into three parts $A_1, A_2$, and $B$ along $L_1$ of lengths $\ell_1, \ell_2$, and $\ell_B$, respectively, as seen in the right picture of Fig.\,\ref{fig:realizations-of-adjacent-geometry}. 
The classical field  $\barphi^\cl_{\nev/2}$ has support on the same torus but cut into two cylinders $A_1\cup A_2$ and $B$. 
Hence,  we have 
\be
\sum_{k=1}^{\nev-1}S[\barphi^\cl_k] &=&
g\hspace{-5pt} \sum_{\substack{ k=1 \\  k\neq \nev/2}}^{\nev-1}
 \int \dif^{\,2} x\, \p_x \barphi^\cl_k\, \p_x \barphi^\cl_k 
  +g  \int \dif^{\,2} x\, \p_x \barphi^\cl_{\nev/2}\, \p_x \barphi^\cl_{\nev/2}\,,  \nn\\ \nn
 %_{\substack{\mu \in\Z^{\nev-2}\, \\  \upsilon \in\Z^{\nev-2}\,\\ \gamma \in \Z}}
% -g L_2 \sum_{k=1}^{\nev-1} \left(\int_{A_1}\dif x +\int_{A_2} \dif x +\int_B \dif x \right)( \p_x f\p_x f)
%\\ \nn
&=& g L_2\hspace{-5pt} \sum_{\substack{ k=1 \\  k\neq \nev/2}}^{\nev-1}\hspace{-5pt} \left({1\over \ell_1} \left( \barphi^\cl_k\vert_{\G_A}-\barphi^\cl_k\vert_{\G_1}\right)^2+ {1\over \ell_2} \left(\barphi^\cl_k\vert_{\G_A}-\barphi^\cl_k\vert_{\G_2}\right)^2+{1\over \ell_B} \left(\barphi^\cl_k\vert_{\G_1}-\barphi^\cl_k\vert_{\G_2}\right)^2\right)\,
\\ 
&&+g L_2  \left({1\over \ell_1+\ell_2}  + {1\over \ell_B}\right)\left( \barphi^\cl_{\nev/2}\vert_{\G_2}-\barphi^\cl_{\nev/2}\vert_{\G_1}\right)^2 .
\ee
Using the boundary conditions \eqref{B-torus-new-notation-v3} it is easy to compute the above expression. We only need to notice that for $i, j=1\,, \dots, \nev/2-1$
\be\label{mix-prod}
\left((M_{\nev/2-1})_{ij} x_j - \sqrt{1-{2\over \nev}} \delta_{i, \nev/2-1} \gamma\,\right)^2 & =&
x^T M_{\nev/2-1}^T M_{\nev/2-1} x - {4\over \nev} \gamma\,  I\cdot x-\left(1-{2\over \nev}\right) \gamma^2\,, 
\nn\\
&=&x^T \,T_{\nev/2-1}\, x - {4\over \nev} \gamma\,  I\cdot x-\left(1-{2\over \nev}\right) \gamma^2 \,,
\ee
since $M_{\nev/2-1}^T M_{\nev/2-1}  = \, :T_{\nev/2-1}$~\cite{Zhou:2016ykv}, and $I=(1\, \dots, 1)\in \Z^{\nev/2-1}$. 
The matrix $ T_{m-1}$  appeared already in~\cite{Zhou:2016ykv,Angel-Ramelli:2019nji}, and it is given by
\be\label{Tmatrix}
 T_{m-1} :  = M^T_{m-1} M_{m-1}= 
 \begin{bmatrix}
1-{1\over m} & -{1\over m } & \dots & -{1\over m } \\
 -{1\over m } & 1-{1\over m}  & \dots & -{1\over m } \\
\vdots 
\\
%~& ~{1\over \sqrt{n(n-1)}} & \dots~ & ~\dots & -\sqrt{1-{1\over n}}\\
-{1\over m }& -{1\over m } & \dots  & 1-{1\over m}\\
\end{bmatrix}.\qquad
\ee
The specific coefficients in the expression \eqref{mix-prod} are simply due to the explicit form of the matrix $M_{\nev/2-1}$, which is obtained by deleting the $\nev/2$-th row and column from $U_{\nev/2}$, given in \eqref{def-matrix-U}. 
Hence, a straightforward computation gives the expression \eqref{we-torus-v1}, where we have collected the modes as
$$
\mu: = (\mu^1, \mu^A)\,, \qquad \upsilon: = (\upsilon^1, \upsilon^A)\,,
$$
and used the aspect ratios
\be
\label{again-aspect-ratios}
u_1= {\ell_1\over L_1}\,, \qquad u_2= {\ell_2\over L_1}\,, \qquad u_B= {\ell_B\over L_1}=1-u_1-u_2 \equiv 1-u_{12}\,. 
\ee

\noindent
Thus,  we obtain the following expression for $\WE$
\be
\label{we-torus-v1}
\WE(\nev)\nn\\\nn
&& \hspace{-35pt}=\sum_{\Z^{2\nev-3}} \exp\left\{- {g (2 \pi R_c)^2 \over \vert\tau\vert}\left[ 
\mu^T \, \mathrm T \, \mu +\upsilon^T \, \mathrm T \, \upsilon
+\left({1-\frac 2n}\right){1-u_2\over u_1 (1-u_{12})}\left(\gamma -I\cdot \mu^1\right)^2 \right.\right.
%+\left({1-\frac 2n}\right)\left({1\over u_1}+{1\over u_B}\right)\left(\gamma -I\cdot \mu^1\right)^2 \right.\right.
\\ \nn
&&%\hspace{-33pt}+\left.\left. \frac 1n\left(\frac{1}{u_B}+\frac{1}{u_1+u_2}\right)\left(\gamma -I \cdot \upsilon^1\right)^2 
\hspace{-33pt}+\left.\left.\frac{1}{n\, u_{12}(1-u_{12})}\left(\gamma -I \cdot \upsilon^1\right)^2 
%-{4\over n } \left(\gamma-I\cdot \mu^1 \right) \left( \left({1\over u_1}+{1\over u_B}\right)I\cdot \mu^1 - {1\over u_1}I\cdot \mu^A \right)\, 
-{4\over n\, u_1 } \left(\gamma-I\cdot \mu^1 \right) \left( {1-u_2\over u_1 (1-u_{12})}I\cdot \mu^1 - I\cdot \mu^A \right)\, 
%-{4\over n u_B} \gamma \left( I\cdot \mu^1\right)
\right]
\right\}, \\
\ee
where the $2\nev-2\times 2\nev -2$ matrix $\mathrm T$ is given by 
\be
\label{t-odd}
&&\mathrm T = {1\over u_1} \begin{bmatrix}
\frac{1-u_2}{1-u_{12}} \, T_{\nev/2-1} & -T_{\nev/2-1} \\
- T_{\nev/2-1} & \frac{u_{12}}{u_2} \, T_{\nev/2-1} \\
\end{bmatrix} \,,
\ee
and $T_{\nev/2-1}$ has been defined above, see \eqref{Tmatrix}. 
%Note that the matrix $ T_{m-1}$ appeared already in~\cite{Zhou:2016ykv,Angel-Ramelli:2019nji}.
The second term in the first line of $\WE$  \eqref{we-torus-v1} sources a coupling between the two sets of frequencies $\mu^1$ and $\upsilon^1$ (corresponding to the odd and even labelled fields originally), and it is responsible for the non-factorisation of the winding sector in the toroidal case. 
Indeed, the cut $\G_1$ carries only one degree of freedom, represented by the cut function $\chi_1$, and thus $\nev-1$ winding  modes, and it does not see the factorisation into two sets of $\nev/2-1$ modes which is present at the cut $\G_A$ due to the partial transposition. 
We can shift the mode $\gamma$ as 
\be
\gamma\to \gamma-I\cdot \mu^1\,,
\ee
since we are summing over all integers, and we obtain 
\be
\label{we-torus-v2}
\WE(\nev)\nn\\\nn
&& \hspace{-35pt}=\sum_{\Z^{2\nev-3}} \exp\left\{- {g (2 \pi R_c)^2 \over \vert\tau\vert}\left[ 
\mu^T \, \mathrm T \, \mu +\upsilon^T \, \mathrm T \, \upsilon
+\left({1-\frac 2n}\right){1-u_2\over u_1(1-u_{12})}\left(\gamma -I\cdot \mu^1\right)^2 \right.\right.
\\ \nn
&&\hspace{-33pt}+\left.\left.{1\over n\, u_{12}(1-u_{12})}\left(\gamma -I \cdot \upsilon^1\right)^2 
-{4\over n\, u_1 } \left(\gamma-I\cdot \mu^1 \right) \left( {1-u_2\over 1-u_{12}} I\cdot \mu^1 - I\cdot \mu^A \right)\, 
%-{4\over n u_B} \gamma \left( I\cdot \mu^1\right)
\right]
\right\}. \\
\ee
It is useful to rewrite $\WE$ as a multi-dimensional theta function. 
Defining $\Omega \in \Z^{2\nev -3}$ as $\Omega= (\mu,\upsilon, \gamma )$,  \eqref{we-torus-v2} can be written as 
\be
\label{we-torus-v4-2}
\WE(\nev) &=& \sum_{\Omega \in \Z^{2\nev-3}} \exp\left(- {g (2 \pi R_c)^2 \over \vert\tau\vert} \Omega^T \, \mathcal{T} \, \Omega\right),
\ee
where the matrix $\mathcal T$
is given by 
\be
\label{huge-T}
\renewcommand{\arraystretch}{1.5} % give some more room
\mathcal T=
\left[\begin{array}{@{}c | c|c@{}}
A_{\nev -2 \times\nev -2 } & \mathbf{0}_{\nev -2 \times\nev -2 } &
  \begin{matrix}
- {1-u_2\over u_1(1-u_{12})}I_{\nev/2-1} \vspace{7pt}\\
 \hdashline\vspace{5pt}

  {2\over n u_1} I_{\nev/2-1}
  \end{matrix}
\\ \hline
  \mathbf{0}_{\nev -2 \times\nev -2 } &  B_{\nev -2 \times\nev -2 }&
  \begin{matrix}
-{1\over n (1-u_{12})u_{12}} I_{\nev/2-1}\vspace{7pt}\\
 \hdashline\vspace{3pt}
 0_{\nev/2-1}
  \end{matrix}
 \\ 
  \hline
  - {1-u_2\over u_1(1-u_{12})} I^T_{\nev/2-1} ~~~  {2\over n u_1} I^T_{\nev/2-1}  &  {-1\over n (1-u_{12}) u_{12}} I^T_{\nev/2-1} ~~~ \mathbf{0}^T_{\nev/2-1} &  \frac{n-1}{n (1-u_{12}) u_{12}}{+}\frac{(n-2)u_2}{n u_1 u_{12}}
\end{array}\right]\nn\\
\ee
and the matrices $A$ and $B$ are defined as 
\be
&&\hspace{-10pt}
A_{\nev -2 \times\nev -2 } = 
 \begin{bmatrix}
 {1-u_2\over u_1(1-u_{12})}T_{\nev/2-1} + (1+{2\over n}) {1-u_2\over u_1(1-u_{12})} \mathcal{I}_{\nev/2-1} ~&~ -{1\over u_1}T_{\nev/2-1} -{2\over n u_1}\mathcal{I}_{\nev/2-1}
 \\
 -{1\over u_1}T_{\nev/2-1} -{2\over n u_1}\mathcal{I}_{\nev/2-1}  ~&~ ({1\over u_1}+{1\over u_2})T_{\nev/2-1} \\
 \end{bmatrix},\nn \\ \nn
 &&
 \\
&&\hspace{-10pt}
B_{\nev -2 \times\nev -2 } =
 \begin{bmatrix}
 {1-u_2\over u_1(1-u_{12})}T_{\nev/2-1} +  \frac{1}{n (1-u_{12}) u_{12}} 
 \mathcal{I}_{\nev/2-1} ~ &~ -{1\over u_1}T_{\nev/2-1}\\
~ -{1\over u_1}T_{\nev/2-1}~&~ ({1\over u_1}+{1\over u_2})T_{\nev/2-1} \\
  \end{bmatrix},
 \ee
%
%\be
%a &=& \left(1+{2\over n}\right) \left( {1\over u_1}+{1\over u_B} \right)\\ \nn
%b &=& \frac{1}{n u_B (u_1+u_2)} 
%\ee
with $ \mathcal{I}_{m}$ an $m\times m$ matrix with all entries equal to $1$, that is
\be
 \mathcal{I}_{m} =
  \begin{bmatrix}
 ~ 1 ~& 1 ~& \dots~ &1 ~\\
  ~1 ~& 1 ~& \dots~ &1~ \\
  \vdots &\vdots & \ddots &\vdots\\
    ~1 ~& 1~ & \dots~ &1~ \\
   \end{bmatrix}. 
\ee
%

%%%%%%%%%%%%%%%%%%%%%%%%%%%%%
\subsection{Winding sector $W_{OE}$}
\label{app-winding-sector-oe}

Here, we report the calculation of the winding sector $W_{OE}$ \eqref{oee-winding-torus-gen} which appears in the study of the odd entropy for a toroidal manifold. 
The classical fields $\tilde \phi^\cl$ are defined on the whole torus $L_1 \times L_2$, with entangling cuts $\G_1, \G_2$, and $\G_A$ specified by the boundary conditions~\eqref{bc-woee-torus}. 
Hence, we can write 
\be\label{class-sum-oe-appendix}
\sum_{k=1}^{\nod-1}S[\tilde \phi^\cl_k] &=&\nn
g\hspace{-5pt} \sum_{k}^{\nod-1}
 \int \dif^{\,2} x\, \p_x \tilde\phi^\cl_k\, \p_x \tilde\phi^\cl_k
 \\ \nn
 &=&
g L_2\hspace{-5pt} \sum_{k=1}^{\nev-1}\hspace{-5pt} \left({1\over \ell_1} \left( \tilde\phi^\cl_k\vert_{\G_A}-\tilde\phi^\cl_k\vert_{\G_1}\right)^2+ {1\over \ell_2} \left(\tilde\phi^\cl_k\vert_{\G_A}-\tilde\phi^\cl_k\vert_{\G_2}\right)^2+{1\over \ell_B} \left(\tilde\phi^\cl_k\vert_{\G_1}-\tilde\phi^\cl_k\vert_{\G_2}\right)^2\right)\,
\\ \nn
 &=&g L_2 (2 \pi R_c)^2\, \left(\left({1\over \ell_1}+{1\over \ell_2} \right) \nu_A^T T_{\nod -1} \nu_A + \left({1\over \ell_1}+{1\over \ell_B} \right) \nu_1^T T_{\nod -1}\nu_1-{2\over \ell_1} \nu_1^T T_{\nod -1} \nu_A\right),\\\hspace{-5pt}
\ee
where in the last line we  used the boundary conditions~\eqref{bc-woee-torus}. 
Defining the vector $\nu=(\nu_1, \nu_A)\in \Z^{2\nod-2}$, and the $2\nod-2 \times 2\nod -2$ matrix $\mathrm{T_{OE}}$ as
\be\label{def-toee}
\mathrm{T_{OE}}= 
\begin{pmatrix}
	\left(\frac{1-u_2}{1-u_{12}}\right)T_{\nod-1} & -T_{\nod-1} \\
	-T_{\nod-1} & \left(1+\frac{u_1}{u_2}\right)T_{\nod-1} \\
\end{pmatrix} , 
%\,, \qquad u_2: ={\ell_2\over L_1}, \quad u_B: ={\ell_B\over L_1}\,. 
\ee
where the aspect ratios were defined in \eqref{again-aspect-ratios},  we can rewrite \eqref{class-sum-oe-appendix} as 
\be
\sum_{k=1}^{\nod-1}S[\tilde \phi^\cl_k] &=& g {(2 \pi R_c)^2 \over \vert \tau\vert u_1} \nu^T \, \mathrm{T_{OE}} \, \nu\,.
\ee
Thus, the explicit expression for the winding sector reads 
\be\label{winding-torus-oee}
W_{OE}(\nod) =\sum_{\phi^\cl_i} e^{-\sum_i S[\phi^\cl_i]}
=\sum_{\vec\nu\in \Z^{2\nod -2}}\exp\left\{- \pi \, {4 \pi\, g\, R^2_c \over |\tau| u_1}\,  \vec\nu^T \, \mathrm{T_{OE}} \, \vec \nu\right\}. 
\ee 
The matrix $\mathrm{T_{OE}}$ is positive definite in the physical region where the aspect ratios $u_1, u_2, u_B$ are positive and constrained  to satisfy $u_1+u_2+u_B=1$, $u_{1, 2, B}<1$. 
This guarantees the convergence of the series, indeed the sum in \eqref{winding-torus-oee} is nothing but a multi-dimensional theta function. 
We can use the reciprocal formula for the multi-dimensional $\theta$ function, cf. \eqref{reciprocal-formula-multi}, 
and write $W_{OE}(\nod)$ as 
\be
W_{OE}(\nod) = 
{\left({4 \pi g R_c^2\over |\tau| u_1} \right)^{-(\nod-1)}\over \det T_{\nod-1}}\hspace{-3pt} \left({u_1\over u_2 (1-u_{12})}\right)^{\hspace{-2pt}-(\nod-1)/2} \hspace{-10pt}
\sum_{\vec \mu \in \Z^{2\nod-2}} %\exp\left\{ -{\pi\over c} \vec\mu \, \hat {\mathrm T}^{-1} \vec \mu \right\}
\hspace{-5pt} \exp\left\{ - \pi \, {|\tau| (1-u_{12}) u_2 \over 4 \pi g \, R_c^2}\,  \vec\mu^T \, \hat {\mathrm T}^{-1}\,  \vec \mu \right\},\nn\hspace{-10pt}\\
\ee
where the inverse matrix is
\be
\hat {\mathrm T}^{-1} = 
\begin{pmatrix}
	\left(1+\frac{u_1}{u_2}\right)T^{-1}_{\nod-1} & T^{-1}_{\nod-1} \\
	T^{-1}_{\nod-1} & \left(\frac{1-u_2}{1-u_{12}}\right)T^{-1}_{\nod-1} \\
\end{pmatrix} 
. 
\ee
The determinant of $T_{\nod-1}$ is simply $1/\nod$ \cite{Zhou:2016ykv}, hence we can write 
\be
W_{OE}(\nod) = 
\nod {\left({ |\tau|\over 4 \pi g R_c^2} \right)^{\hspace{-2pt}(\nod-1)}}\hspace{-4pt} \left({ u_1 u_2 (1-u_{12})}\right)^{(\nod-1)/2} \hspace{-10pt}
\sum_{\vec \mu \in \Z^{2\nod-2}} %\exp\left\{ -{\pi\over c} \vec\mu \, \hat {\mathrm T}^{-1} \vec \mu \right\}
\hspace{-8pt} \exp\left\{ - \pi \, {|\tau| (1-u_{12}) u_2 \over 4 \pi g \, R_c^2}\,  \vec\mu^T \, \hat {\mathrm T}^{-1}\,  \vec \mu \right\}.\nn\hspace{-10pt}\\
\ee
The eigenvalues of $\hat {\mathrm T}^{-1}$ are straightforward to compute, and we find 
\be
&& \{\nod \, \lambda_+\, , \nod\,  \lambda_-\, , \underbrace{\lambda_+\, , \dots\,, \lambda_+}_{\nod -2}\,,  \underbrace{\lambda_-\, , \dots\,, \lambda_-}_{\nod -2}\, \}\,, 
\ee
with 
\be\nn
\lambda_\pm %&=& 1+{u_1\over 2} \left({1\over u_B}+{1\over u_2}\right) \pm \sqrt{ 1+ {u^2_1\over 4} \left({1\over u_B}-{1\over u_2}\right)^2  }\,\\\nn
&=&
1+{u_1\over 2 u_2} {1-u_1\over 1-u_{12}} \pm \sqrt{ 1+ {u^2_1\over 4} \left({1\over 1-u_{12}}-{1\over u_2}\right)^2  }\,. 
\ee
Using the multi-dimensional theta function we can write the winding sector as
\be\label{winding-torus-oee-final}
W_{OE}(\nod) = 
\nod \, {\left({ |\tau|\over 4 \pi g R_c^2} \right)^{(\nod-1)}} \left({ u_1 u_2 (1-u_{12})}\right)^{(\nod-1)/2} 
\Theta(\vec 0\,\vert \,\mathrm{U})\,,
\ee
where $\Theta$ and $\mathrm U$ are given by
\begin{gather}
\label{eq:appendix-winding-OE-final}
\Theta(\vec 0\,\vert \,\mathrm{U}) = \sum_{\vec \mu \in \Z^{2\nod-2}}e^{- \pi \,  \vec\mu^T \, \mathrm U\,  \vec \mu}\,,\\
\mathrm{U}= {|\tau| (1-u_{12})u_2 \over 4 \pi g \, R_c^2} R^T
\begin{pmatrix}
	\nod \, \lambda_+& & & \\
	& \nod \, \lambda_+ & &\\
	&&  \lambda_+ &\\
	& & & \ddots & \\
	& & & & \lambda_-
\end{pmatrix} 
R\,,\nn
\end{gather}
and where $R$ is a unitary matrix. 
%

%%%%%%%%%%%%%%%%%%%%%%%%%%%%%%%%%%%%%%%
%%%%%%%%%%%%%%%%%%%%%%%%%%%%%%%%%%%%%%%
\section{Pure state limit for the odd entropy}
\label{app:OE}

In this section we illustrate another example for the pure state limit of the odd entropy.
Since we do not have an analytical continuation for the winding sector $W_{OE}$ of the torus, we can only consider the pure state limit for non-compact fields. 
This is equivalent to considering only the contribution from the partition functions. 
The partition function for non-compact fields on the complete torus is divergent due to the zero mode, for this reason we will place the theory on a cylinder with Dirichlet boundary conditions at its endpoints. 
The contribution to the odd entropy from the fluctuations is essentially given by the expression \eqref{oee-torus-det-explicit}, which now reads 
\be
-\log{ \left({Z_{A_1} Z_{A_2} Z_{B}\over Z_{A\cup B}}\right)}&=& \half \log{{\det \Delta_{\rm cyl, A_1} \det \Delta_{\rm cyl, A_2} \det \Delta_{\rm cyl, B} \over \det \Delta_{\rm cyl}}} \qquad  \\\nn
 &=& \half \log \left({4 u_1 u_2 (1-u_{12}) \vert \tau\vert^2}\right)+ \log \left\vert{{\eta(2 u_1\tau ) \eta(2 u_2\tau ) \eta(2 (1-u_{12})\tau )\over \eta(2\tau )}}\right\vert\,,
 \ee
 where in the last step we used the results \eqref{det-cyl} collected in Appendix \ref{app:det-torus}. 
In the limit $u_B=1-u_{12} =\varepsilon \to 0$, the above expression becomes
\be
%\half \log{{\det \Delta_{\rm cyl, A_1} \det \Delta_{\rm cyl, A_2} \det \Delta_{\rm cyl, B} \over \det \Delta_{\rm torus}}} +\half \log{4 \pi g R_c^2 \mathcal A} \qquad  \\\nn
 %&=& 
% \half \log \left({8 u_1 u_2 u_B \vert \tau\vert^2}\right)+ \log {{\eta(2 u_1\tau ) \eta(2 u_2\tau ) \eta(2 u_B\tau )\over \eta^2(\tau )}}+\half \log{4 \pi g R_c^2}\,,
-\log{ \left({Z_{A_1} Z_{A_2} Z_{B}\over Z_{A\cup B}}\right)}\approx
\half \log \left( 2\vert \tau\vert u_1 (1-u_1)\right) +\log\left\vert{{\eta(2 u_1\tau ) \eta(2 (1-u_1)\tau )\over \eta(2\tau )}}\right\vert -{\pi\over 24\varepsilon \vert\tau\vert}+ \dots \,.\;\nn\\
 \ee
As explained in the main body, see discussions around \eqref{OE-pure-limit-reg}, we need to consider a ``regulated'' odd entropy, where the contributions from the entangling surface at $\G_2$ are correctly subtracted. 
Our choice is to use the  entanglement entropy for the corresponding density matrix $\rho_A$. 
Concretely, it means that we subtract the universal term of the entanglement entropy of a bipartite cylinder, whose expression was initially obtained in~\cite{Zaletel2011, Zhou:2016ykv} (here we only need the non-compact contribution):
\be
S_{EE}(\rho_A)= \half \log(2u(1-u)\vert\tau\vert)+\log\left\vert{\eta(2u\tau) \eta(2(1-u)\tau)\over \eta(2\tau)}\right\vert\,,
\ee
where here $u=u_1+u_2$. 
When $u_B\to 0$ (i.e. $u_1+u_2\to 1$), the above expression becomes
\be
S_{EE}(\rho_A) \approx -{\pi\over 24\varepsilon \vert\tau\vert} \,.
\ee
 Hence, in  the limit $u_B\to 0$, the ``regulated'' odd entropy is
 \be
 \Delta S_o \approx\half \log \left( 2\vert \tau\vert u_1 (1-u_1)\right) +\log\left\vert{{\eta(2 u_1\tau ) \eta(2 (1-u_1)\tau )\over \eta(2\tau )}}\right\vert\,.
\ee
This is the universal part of the entanglement entropy for a system on a bipartite cylinder, as expected \cite{Tamaoka:2018ned}.

%%%%%%%%%%%%%%%%%%%%%%%%%%%%%%%%%%%%%%%
%%%%%%%%%%%%%%%%%%%%%%%%%%%%%%%%%%%%%%%

%\bibliographystyle{apalike}
%\bibliographystyle{utphys}
%\bibliography{LN-of-the-QLM}

\providecommand{\href}[2]{#2}

\end{document}